\documentclass[5p,times]{elsarticle-own}

\usepackage{amssymb}
\usepackage[table,xcdraw]{xcolor}
\usepackage{mathtools}
\usepackage{hyperref}
\usepackage{microtype}
\usepackage{subcaption}
\usepackage[export]{adjustbox}
\usepackage[normalem]{ulem}
\journal{Future Generation Computer Systems}

\begin{document}

\begin{frontmatter}

\title{Lotaru: Locally Predicting Workflow Task Runtimes for \\ Resource Management on Heterogeneous Infrastructures}

\author[label1]{Jonathan~Bader}\corref{cor1}
\ead{jonathan.bader@tu-berlin.de}
\author[label2]{Fabian~Lehmann}
\ead{fabian.lehmann@informatik.hu-berlin.de}
\author[label3]{Lauritz~Thamsen}
\ead{lauritz.thamsen@glasgow.ac.uk}
\author[label2]{Ulf~Leser}
\ead{leser@informatik.hu-berlin.de}
\author[label1]{Odej~Kao}
\ead{odej.kao@tu-berlin.de}

\cortext[cor1]{Corresponding author}
        
\affiliation[label1]{organization={Technische Universität Berlin},
             country={Germany}}
\affiliation[label2]{organization={Humboldt-Universität zu Berlin},
             country={Germany}}
\affiliation[label3]{organization={University of Glasgow},
             country={United Kingdom}}

\begin{abstract}
Many resource management techniques for task scheduling, energy and carbon efficiency, and cost optimization in workflows rely on a-priori task runtime knowledge.
Building runtime prediction models on historical data is often not feasible in practice as workflows, their input data, and the cluster infrastructure change.
Online methods, on the other hand, which estimate task runtimes on specific machines while the workflow is running, have to cope with a lack of measurements during start-up.
Frequently, scientific workflows are executed on heterogeneous infrastructures consisting of machines with different CPU, I/O, and memory configurations, further complicating predicting runtimes due to different task runtimes on different machine types.

This paper presents Lotaru, a method for locally predicting the runtimes of scientific workflow tasks before they are executed on heterogeneous compute clusters.
Crucially, our approach does not rely on historical data and copes with a lack of training data during the start-up.
To this end, we use microbenchmarks, reduce the input data to quickly profile the workflow locally, and predict a task's runtime with a Bayesian linear regression based on the gathered data points from the local workflow execution and the microbenchmarks.
Due to its Bayesian approach, Lotaru provides uncertainty estimates that can be used for advanced scheduling methods on distributed cluster infrastructures. 

In our evaluation with five real-world scientific workflows, our method outperforms two state-of-the-art runtime prediction baselines and decreases the absolute prediction error by more than 12.5\%.
In a second set of experiments, the prediction performance of our method, using the predicted runtimes for state-of-the-art scheduling, carbon reduction, and cost prediction, enables results close to those achieved with perfect prior knowledge of runtimes.

\end{abstract}


\end{frontmatter}

\section{Introduction}
\label{sec:introduction}
Scientists from many domains have to analyze an increasing amount of data to approach their research goals~\cite{berriman2004montage,Sudmanns_Tiede_Augustin_Lang_2019,yates2021reproducible,garcia2020sarek,muirRealCostSequencing2016,schaarschmidt2021workflow,stein2019progress}.
In remote sensing, scientists analyze many high resolutions images from long satellite missions~\cite{berriman2004montage, Sudmanns_Tiede_Augustin_Lang_2019}, in genomics, scientists analyze many read sets~\cite{yates2021reproducible,garcia2020sarek, muirRealCostSequencing2016}, in material science, scientists analyze many molecules~\cite{schaarschmidt2021workflow,stein2019progress}. 
Scientific workflow management systems (SWMS) help scientists to compose, execute, and monitor their analysis workflows over large data sets on distributed infrastructures~\cite{deelman2019evolution, witt2019feedback, nextflow}.

\begin{figure}[th]
    \includegraphics[width=\columnwidth]{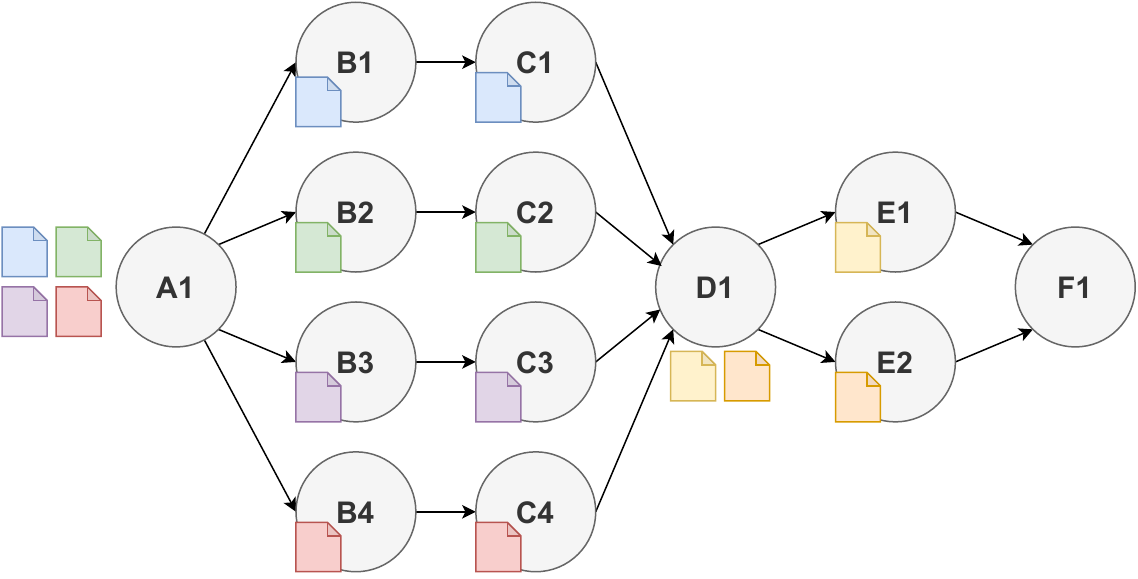}
    \caption{Example execution of a scientific workflow.}
	\label{fig:example_workflow}
\end{figure}

Many SWMS handle a workflow as a directed acyclic graph (DAG), consisting of a set of tasks T and a set of directed edges E.
A task is a wrapper for an application, executed as an atomic unit that transforms input data to output data. 
An edge describes the data flow, defines a dependency between two tasks, and thus constrains the execution order.
As illustration, Figure~\ref{fig:example_workflow} depicts the execution of an example workflow with a single starting task, Task A1, that checks the input.
Then, four instances of Task B are executed in parallel, each operating on a single input file and each resulting in the execution of a Task C instance.
Task D1 combines the results from the preceding task executions, while two instances of Task E operate on the output of Task D1 in parallel.
Task F1 serves as the sink task and merges the results.

Scientific workflows are frequently executed over large amounts of data, leading to huge workflow graphs and, thus, long runtimes that can easily exceed days or weeks~\cite{cybershake,ferreira2019accurately, da2020characterizing}.
Hence, from a resource management perspective, makespan minimization~\cite{heft,pheft}, carbon emission reduction~\cite{wiesner2021let,carbonAware}, or predicting and optimizing the cost of running the workflow are important optimization objectives~\cite{rosa2018cost,rosa2021computational}.
Many state-of-the-art methods for these objectives rely on accurate predictions of resource requirements and the runtime of each task~\citep{dziok2016adaptive}.
For instance, the state-of-the-art scheduling heuristic \emph{HEFT} requires accurate runtime estimates for each task-node pair~\cite{heft}.
The carbon reduction approach \emph{Let’s Wait Awhile} assumes exact task runtimes to temporally shift the workflow to periods with lower carbon emissions~\cite{wiesner2021let}.
The \emph{Computational Resource and Cost Prediction service} (CRCPs) reports the cost of workflow execution based on predicted runtimes~\cite{rosa2021computational}. 

In practice, predicting a task’s runtime for a single machine is not sufficient since infrastructures available to scientists are frequently heterogeneous,
consisting of machines with different CPU, I/O, and memory configurations.
The reasons include partially upgraded nodes, hardware replacements over time, or clusters with different machines from the beginning~\cite{cpuperformance,hutson2019managing}.
In such heterogeneous settings, not only the number of cores, the amount of memory, or the disk size differs.
For instance, different processors have different clock rates and cache sizes, memory can differ regarding latency and frequency, and disks have different read/write capabilities.  
In cloud environments such as Amazon AWS, Microsoft Azure, or Google Cloud, cloud providers offer machine instances from different manufacturers with various processor types, memory configurations, or storage solutions~\cite{schad2010runtime}, leading to an inherent heterogeneous landscape.  
Due to the heterogeneity aspect of infrastructures, the same task will yield different runtimes, consume different amounts of resources, and produce varying amounts of carbon emissions on different nodes.
This opens the problem that resource management components need accurate predictions not only per task but actually per task-node pair.
Such information, however, is very often unavailable~\cite{schwiegelshohnHowDesignJob2015}.

In our paper, we address this problem by locally predicting task runtimes before the start of a specific workflow execution on a heterogeneous cluster infrastructure.
Lotaru uses microbenchmarks on the target infrastructure, local workflow executions on downsampled partitions of the entire input, and a Bayesian regression method to predict task runtimes based on the gathered data.
Our method is designed to predict the runtime of scientific workflow tasks executed on heterogeneous clusters by profiling them on a scientist's personal computer, aiming to avoid the use of often scarce cluster resources, thereby maximizing system efficiency.
It is intended for workflows for embarrassingly parallel problems where the same \mbox{(sub-)workflows} are executed over many inputs or intermediate results.

Our method does not depend on any historical information but performs all measurements and predictions before the start of a specific workflow execution.
Notably, this also allows for scenarios where the learned models are reused for future executions of the same workflow over different input data.

The contributions of this paper are:

\begin{itemize}
    \item We propose Lotaru, a method that predicts the runtime of scientific workflow tasks on a scientist's local machine before the workflow is executed on a heterogeneous cluster infrastructure. To this end, we use microbenchmarks, reduce the workflow's input data to quickly execute the workflow locally, and predict the runtime for target nodes with a Bayesian linear regression.
    \item We provide an open-source implementation of our prediction method\footnote{github.com/CRC-FONDA/Lotaru} with an extendable interface for use with other domains and publish traces detailing more than 10,000 task executions on heterogeneous cluster nodes\footnote{github.com/CRC-FONDA/Lotaru-traces}.
    \item We evaluate our method with five real-world workflows using multiple data inputs on a heterogeneous cluster of six machine types.
    Our experiments show that our local task runtime prediction method outperforms the baselines regarding prediction error. 
    Further, we used the resulting predictions for several resource management techniques, showing results close to using accurate runtime predictions. 
\end{itemize}

\begin{figure*}[ht!]
    \includegraphics[width=\textwidth]{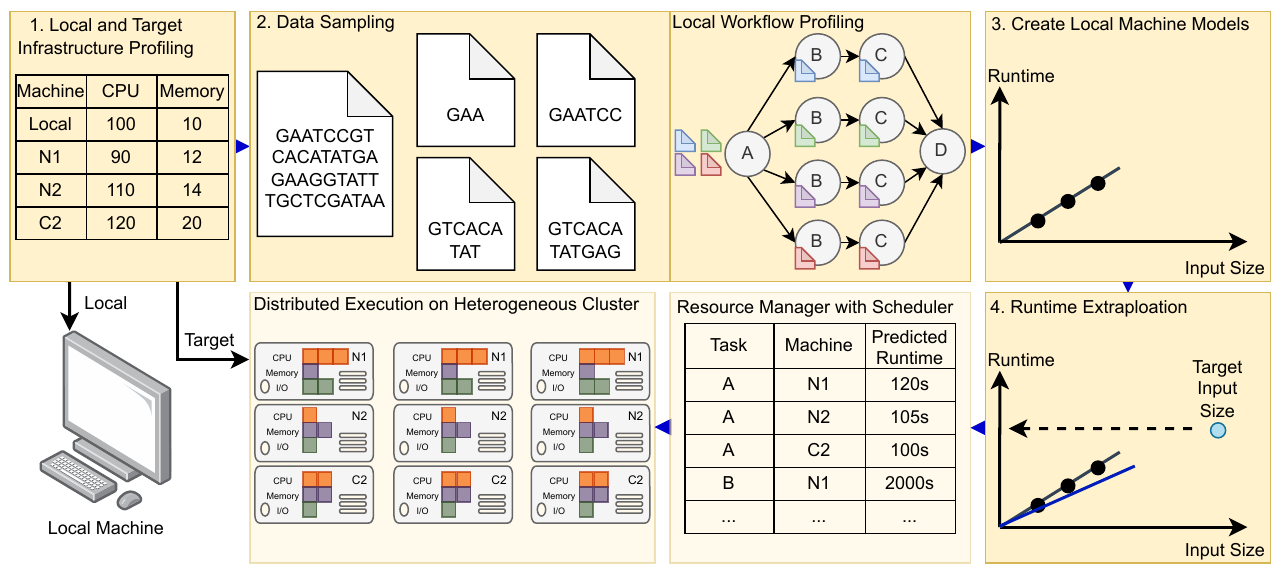}
    \caption{Lotaru profiles the local and target infrastructure with microbenchmarks, reduces the input data to quickly execute the workflow locally on a single machine, and predicts the runtime for target cluster nodes with a Bayesian linear regression based on the gathered data points from the local workflow profiling and the microbenchmarks. The cluster's resource manager's scheduling unit can then use these predictions to execute the workflow tasks on the compute nodes in a distributed manner.} \label{fig:arch}
\end{figure*}

\section{Lotaru Overview}
\label{sec:overview}
Figure~\ref{fig:arch} provides an overview of our approach and the execution environment.
The steps highlighted in dark yellow are part of our methodology, and the steps highlighted in transparent yellow refer to the execution environment, that is, using our predictions to run workflow tasks on the cluster infrastructure.

In the first step, the infrastructure profiler analyzes the performance characteristics of a local computer (e.g., the scientist's computer) and the target infrastructure nodes (e.g., a heterogeneous commodity cluster).
Infrastructure microbenchmarks are used to quickly measure CPU, memory, and I/O on the local computer and the different target nodes. 
The benchmarks can be extended by application-specific benchmarks for certain well-known tasks, e.g., FastQC or BWA in the genomics domain. 

Next, input data for the workflow profiling on the local machine is required.
This data can either be automatically generated by selecting one of the foreseen input files and downsampling or slicing it into several smaller files or can be provided manually by the scientist, e.g., a small genome file or a small satellite image.
Then, the workflow is executed on the local machine with the previously generated small inputs to profile task characteristics.
During this execution, monitoring data, such as task runtimes, input sizes, and read/write I/O, are collected.

Next, we train a Bayesian linear regression model for each task using the measurements from running the workflow locally.
Our model assumes a relationship between a task's input size and its runtime, using the input size as the independent variable.
This model can then be used to predict task runtimes for arbitrary task input sizes. 
The Bayesian approach also provides lower and upper uncertainty bounds at different confidence levels, expressing that the point estimate is likely to be inaccurate.

In step four, we extrapolate the predicted runtimes to the nodes on the target cluster infrastructure.
To this end, we leverage the nodes' microbenchmark measurements and set them in relation to the local microbenchmark measurements.
In the case of application-specific benchmarks, metrics from
these specific task benchmarks serve as the basis for extrapolating the runtime as they are more characteristic of the respective task.

Lastly, resource managers, such as Kubernetes or Slurm, can incorporate these predictions to effectively execute workflow tasks across a distributed infrastructure.

\section{Related Work}
In this section, we focus on runtime prediction methods in general, followed by task-runtime prediction approaches for heterogeneous infrastructures and resource management methods where such runtime predictions are required.
Lastly, we discuss the differences from our own previous work.

\subsection{Task Runtime Prediction}\label{subsec:approaches-for-runtime-prediction-based-on-historical-runtime-data}%
There has been extensive research on workflow task runtime prediction methods.
Many approaches leverage machine learning techniques to solve this problem~\cite{sadjadi2008modeling, da2013toward,da2015online, nadeem2017modeling}.

Nadeem et al.~\cite{nadeem2017modeling} model and predict task execution times on grid environments with the use of neural networks.
Their learning model considers four types of information; a) the workflow structure, e.g., the workflow name or the dependency flow; b) the application, e.g., executable names or files sizes; c) the execution environments, e.g., grid sites or the time of submission, and d) the resource state, e.g., the number of jobs in queue or the jobs running. 
Their study shows which features impact the predicted runtime most and which can be omitted.

Da~Silva~et~al.~\cite{da2013toward, da2015online} predict task resource consumption, such as runtime, disk space, and memory consumption, for tasks in scientific workflows.
Based on monitoring tools and historical data, they apply a regression tree for resource prediction.
Beforehand, they identify data subsets with a high correlation by using density-based clustering.
Then, they predict the expected resource usage for correlated data points based on the ratio in this specific cluster.
For uncorrelated data, the authors test a Normal and a Gamma distribution to generate an estimation value.
As both methods use statistical approaches that do not rely on sufficient historical data being available, we use them as competitors for our method and describe them in more detail in Section~\ref{baselines}.

Offline predictors are generally not applicable for new workflows with new tasks, as for these, no historical traces for model learning are available.
Online predictors have to cope with a lack of runs during start-up~\cite{witt2019feedback, witt2019predictive}.

In contrast, Lotaru is designed as a method that can be applied out-of-the-box for workflows without any historical traces or workflows with changed parameters and configurations on any kind of cluster.

\subsection{Task-Runtime Prediction for Heterogeneous Infrastructures}
\label{subsec:cross-infrastructure}

Since real-world infrastructures often consist of nodes with heterogeneous hardware characteristics, prediction models have to take into account the challenge of generalizing their models across different node types.

Pham~et~al.~\cite{pham2017predicting} predict task runtimes in cloud environments using a two-stage approach. 
Their prediction model distinguishes between pre-runtime parameters, e.g., workflow input data or VM types, and runtime parameters, such as CPU, memory, I/O operations, and bandwidth. 
In the first stage, pre-runtime parameters are considered to derive the runtime parameters for the execution.
In the second stage, the task execution time on a target VM is predicted with a regression model using the output data from the first stage, the workflow input data, and the VM specifications.

Hilman~et~al.~\cite{hilman2018task} use an online incremental learning approach with long short-term memory networks (LSTMs) to predict task runtimes in cloud environments.
Again, the authors define pre-runtime and runtime metrics. 
The pre-runtime metrics consist of task information, VM type, and submission time features, and the runtime metrics of CPU, memory, and I/O features.
The runtime metrics are historical time-series data that are extended during execution and serve as the input for training and updating the model after the task finishes its execution.

Matsunaga~et~al.~\cite{matsunaga2010use} evaluate several machine learning approaches for their capability to predict the task runtime.
The evaluation includes the impact of single features on the prediction accuracy and argues for including as many features as possible, letting the respective algorithm decide on the specific selection. 
Further, they showed that different algorithms perform better for different setups, i.e., an algorithm's efficacy depends on the task and its training data.

Knowledge of exact task runtime is also important for workflow simulation.
WRENCH~\cite{casanova2018wrench} is a workflow simulation tool that models a task with a fixed number of instructions, a number of min/max cores, and memory requirements. 
A machine's hardware is described by processor speed, disk read/write capabilities, and memory. 
The task's runtime is then defined by calculating the number of task operations per second divided by the processor speed. 
WRENCH treats all of these values as known and accurate. 
Therefore, WRENCH is a special case because the runtime is not predicted but rather extrapolated.

Most recent prediction approaches use machine learning methods like neural networks, which are known to require large training data sets to perform well.
In contrast, we use a Bayesian Linear Regression model that already works with few training points and provides uncertainty estimates for its predictions~\cite{mcneish2016using, lee2004evaluation}.
Compared to the related work, we do not include hardware characteristics as features for the runtime prediction since this would increase the necessary training data, especially for the target machines, e.g., traces of running the workflow on the target infrastructures.
Instead, we either apply general microbenchmarks or, for well-known tasks, application-specific microbenchmarks to obtain node-specific characteristics implicitly.

\subsection{Resource Management Relying on Task Runtimes}
Several workflow resource management techniques such as makespan minimization, energy and carbon emission reduction, or cost optimization and prediction rely on a-priori task runtime knowledge to work effectively.

\paragraph{Makespan Minimization Scheduling}
\label{subsec:rw-scheduling}

Despite for their lack of adoption into real workflow systems, the literature on makespan minimization scheduling is extensive.
Existing methods approach makespan minimization by either scheduling statically or dynamically~\cite{dubey2018modified, wang2016hsip}.
Static scheduling refers to assigning tasks to resources in advance, i.e., before workflow execution.
Therefore, these approaches cannot adapt to infrastructure failures or changes in the workflow execution plan.
A prominent representative is the list heuristic HEFT (Heterogeneous Earliest-Finish-Time)~\cite{heft}.
Many extensions of this static heuristic exist, e.g., HCPPEFT~\citep{dai2014synthesized}, P-HEFT~\cite{pheft}, AHEFT~\citep{yuAdaptiveReschedulingStrategy2007}, or DQ-HEFT~\citep{kaurDeepQLearningbasedHeterogeneous2022}.
Some of them, e.g., P-HEFT~\cite{pheft} or AHEFT~\cite{yuAdaptiveReschedulingStrategy2007} are dynamic scheduling approaches and assign tasks to resources during the workflow execution.
Consequently, they can also be applied when the exact physical workflow graph is not known before execution, e.g., the task graph depends on intermediary results~\citep{bux2017hi}.

Crucially, static and dynamic approaches have in common that they require comprehensive knowledge about the execution times of all tasks on all available nodes~\citep{schwiegelshohnHowDesignJob2015}.
However, such execution times are usually not available in advance but must be determined either by asking users for estimates~\citep{schwiegelshohnHowDesignJob2015, ilyushkin2018impact,feitelson2015workload, hirales2012multiple}, by analyzing historical traces~\citep{sadjadi2008modeling,nadeem2017modeling,pham2017predicting, hilman2018task,matsunaga2010use}, or by using some form of online learning~\citep{witt2019feedback,da2015online, witt2019predictive}.
Our method aims to predict the runtime for all task-node pairs in heterogeneous infrastructures to enable the makespan minimization of theoretical scheduling methods in real-world systems.

\paragraph{Energy and Carbon Reduction}

Many techniques aim to increase the energy efficiency or to reduce the carbon emissions.

Warade~et~al.~\cite{warade2022gecost} discuss requirements for energy-aware workflow scheduling and present scheduler optimizations for energy efficiency.
The authors propose several techniques, such as switching off nodes, provisioning additional nodes, or optimizing the CPU frequency.
Such approaches heavily rely on accurate runtime predictions since the scheduler needs to ensure that resources are provisioned or revoked at the correct time.
Inaccuracies would extend the execution time and increase energy consumption.

More specific approaches apply sophisticated scheduling on a workflow level~\cite{garg2021energy,fan2022energy}.
Fan et~al.~\cite{fan2022energy} propose an online two-phase scheduling algorithm that minimizes the energy consumption of tasks.
In the preprocessing phase, the task starting time which relies on the task runtimes is required to generate a scheduling queue with priorities.
The consecutive task scheduling phase also relies on task runtimes, applying sub-deadline-initialization.

Other approaches delay or shift the execution to save energy or carbon emissions~\cite{wiesner2021let,carbonAware, versluis2022taskflow}.
TaskFlow~\cite{versluis2022taskflow} identifies slack time in a workflow execution to either use dynamic voltage and frequency scaling and delay the execution or shift the task to a slower but more energy-efficient node.
CICS~\cite{carbonAware}, Ecovisor~\citep{souza2023ecovisor}, and Let's Wait Awhile~\cite{wiesner2021let} propose a similar idea of shifting workloads, such as batch data processing, to either datacenters with a smaller carbon footprint at a certain time or by delaying the execution to times when the grid's energy mix is associated with fewer carbon emissions.

\paragraph{Cost Prediction and Optimization}

Some resource management methods focus on cost prediction and optimization of workflow executions.

Rosa~et~al.\cite{rosa2018cost,rosa2021computational} propose provisioning services for cloud federations that are able to report the cost of workflow execution beforehand.
This is achieved by predicting execution times with a multiple linear regression model based on historical data.

Alkhanak et al.~\cite{alkhanak2016cost} survey scientific workflow scheduling approaches that focus on cost optimization and give a classification of cost-based metrics.
The authors state that the cost of executing the tasks and the time it takes are inversely proportional but are contradictorily optimizations.
Therefore, the scheduler needs to balance between cost of execution and the workflow time while both objectives are influenced by delays that wrong predictions can cause. 

Again, Alkhanak et al.~\cite{alkhanak2015cost} provide challenges and a taxonomy for cost-aware workflow scheduling.
In their paper, they list the task estimation process as a challenge for cost-aware workflow scheduling.

\subsection{Comparison with Own Previous Work}

We first presented the idea of local task runtime estimation for scientific workflows at the 34th International Conference on Scientific and Statistical Database Management (SSDBM 2022)~\cite{baderLotaruLocallyEstimating2022}. 

In this extended journal article, we improve our methodology by using application-specific benchmarks for more accurate prediction results. 
We show that workflows from the same domain frequently share a set of common tasks, opening the space for such reusable application-specific benchmarks.
We add experiments with much larger data sets (from 154GB to 446GB) to show the robustness of our method.
Further, we conducted three new experiments which provide insight into how Lotaru enables the use of resource management techniques, specifically scheduling, carbon efficiency, and cost prediction.
We extended related work with various resource management techniques and show their dependency on accurate runtime predictions. 
We added a new motivation for our work, showing the need for accurate runtime predictions of several resource management techniques for workflows.
Additionally, we extended and improved the prototype implementation.

\section{Approach}
\label{sec:approach}

Lotaru predicts workflow task runtimes for all nodes in heterogeneous clusters in advance of the actual workflow execution without relying on historical data.

\subsection{Overview}
\label{subsec:approach_overview}
In Figure~\ref{fig:arch}, we provide an overview of the phases in our approach, which are explained in detail in the following sections.
In phase one, our method gathers performance insights about the target infrastructure and the local scientist's machine.
Therefore, we either use general or application-specific microbenchmarks. 
In the second phase, the input for running the workflow locally needs to be obtained.
We select one of the data inputs and create several small input partitions.
The scientist can omit this step by providing small input partitions himself.
Next, the workflow is locally executed with the small partitions obtained from the previous step.
In the third phase, we use the collected data points to build a Bayesian linear regression model that predicts the runtime for the task with arbitrary input sizes on the local machine.
In the last phase, the prediction results need to be extrapolated to fit the target infrastructure.
Here, the profiling results are used to extrapolate the predicted runtimes for each node.
Lastly, a resource management unit can use the predictions.

\subsection{Assumptions}
We make the following five assumptions that limit the scope of applicability:

\begin{enumerate}
    \item[A1:] The workflow execution model presented in Section~\ref{sec:introduction} is used.
	\item[A2:] The workflow has multiple at least partly independent input files such as genome sequences, satellite images, or molecule structures that lead to data-parallel task executions.
	\item[A3:] The workflow's input data can be downsampled or sliced into smaller partitions.
    \item [A4:] The workflow is still executable on the downsampled data or sliced partitions.
    \item [A5:] A task's runtime increases linearly with increasing input data size.
    
\end{enumerate}

Assumption A3 requires specific tools to automatically downsample the input. 
These tools vary depending on the data inputs used and are difficult to generalize due to the potentially arbitrary workflow inputs.

In cases where A2 is fulfilled but A3 is not, there are two alternatives to make our method work. First, the scientist manually provides smaller inputs for running the workflow locally, e.g., smaller genome sequences or satellite images with lower resolutions. Second, a subset from the multiple input files can be used at the price of a longer local workflow execution.

When neither A2 nor A3 is fulfilled, the workflow is sequentially executed and consists of data that can not be downsampled. In such cases, the scientist could still provide smaller inputs and the workflow could run multiple times with each of these. However, in such scenarios, it is questionable if the scientist can provide such inputs and if the effort justifies the outcome since sequentially executed tasks require less optimization.

A4 addresses the workflow's executability with the downsampled data to enable a quick local workflow execution.
For instance, certain workflow tasks may rely on the completeness of a data set or necessitate a minimum amount of data, which, if not met, can result in failed task executions.
When A4 is violated, only a subset of the workflow tasks can be profiled locally, resulting in a prediction for only the subset of tasks.

\begin{figure}
\centering
\begin{subfigure}{.23\textwidth}
   \centering
  \includegraphics[width=1\textwidth]{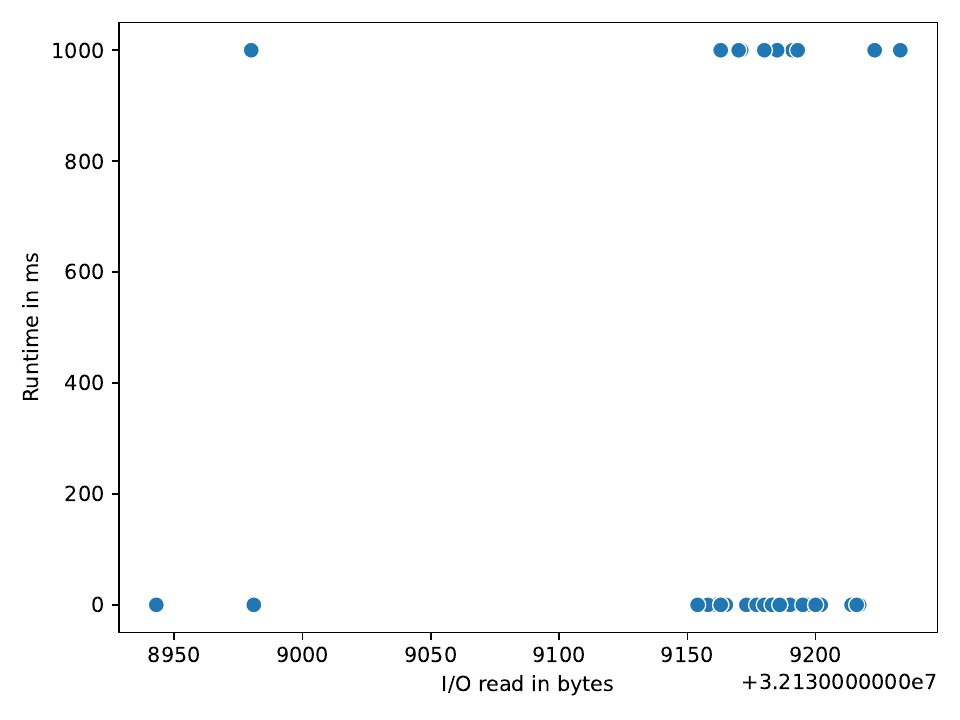}
  \caption{MultiQC}
  \label{fig:sub-first}
\end{subfigure}
\begin{subfigure}{.23\textwidth}
  \centering
  \includegraphics[width=1\textwidth]{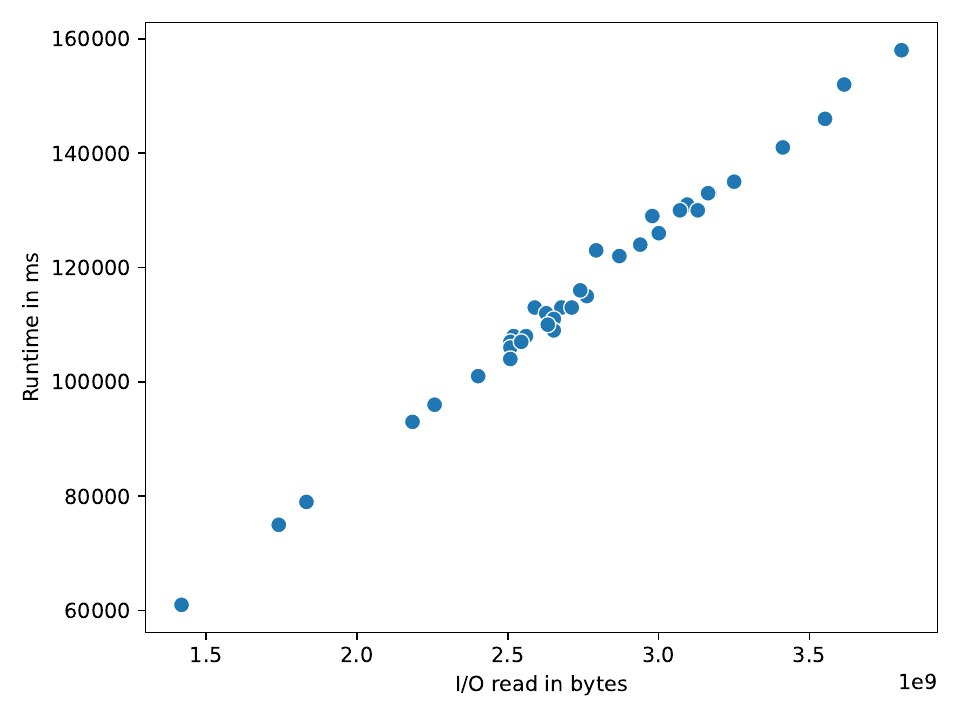}
  \caption{FastQC}
  \label{fig:sub-second}
\end{subfigure}
\begin{subfigure}{.23\textwidth}
   \centering
  \includegraphics[width=1\textwidth]{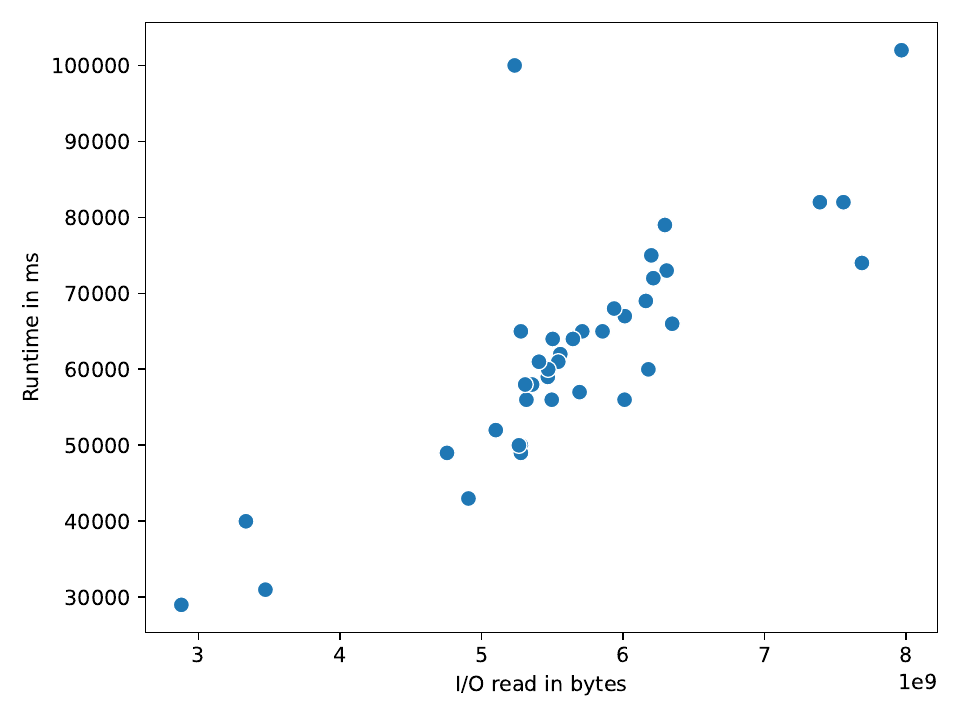}
  \caption{Samtools}
  \label{fig:sub-third}
\end{subfigure}
\begin{subfigure}{.23\textwidth}
  \centering
  \includegraphics[width=1\textwidth]{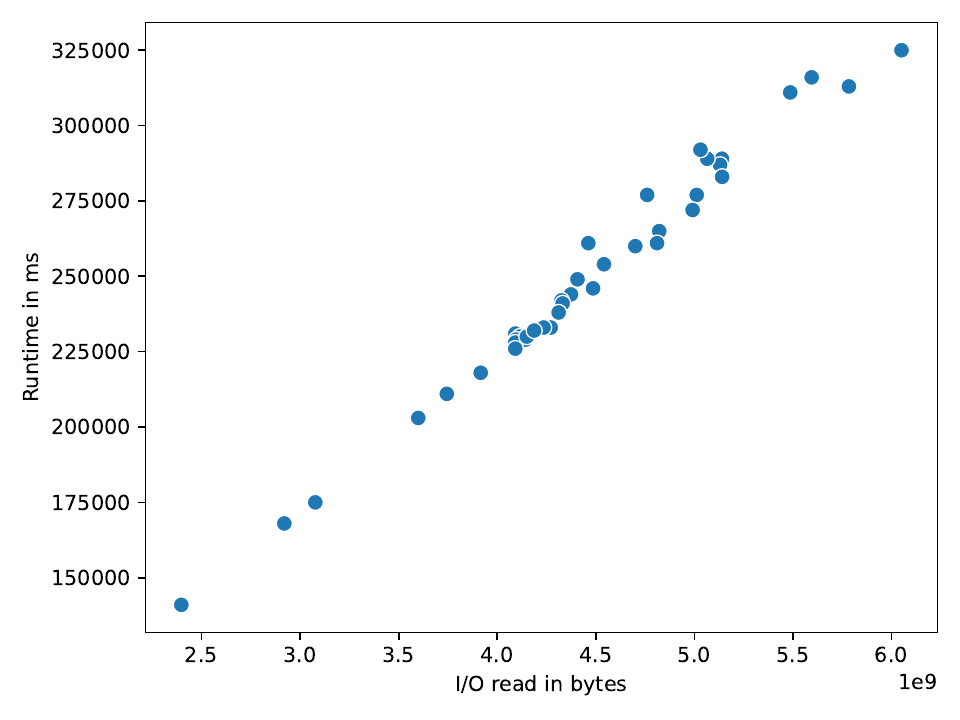}
  \caption{Bedtools}
  \label{fig:sub-fourth}
\end{subfigure}
\begin{subfigure}{.23\textwidth}
   \centering
  \includegraphics[width=1\textwidth]{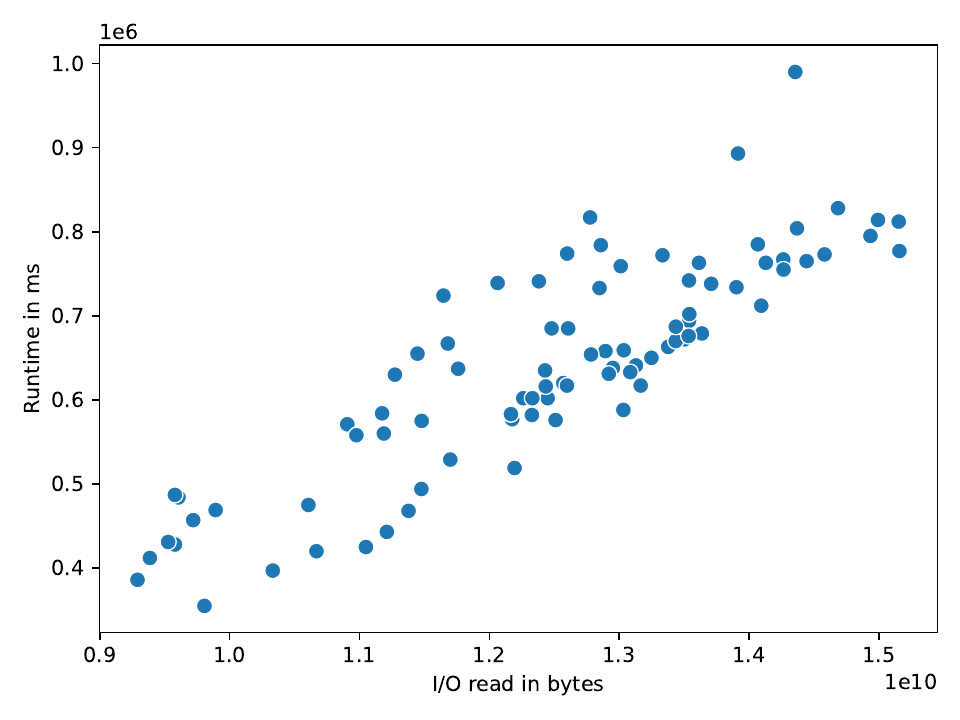}
  \caption{BWA}
  \label{fig:sub-fith}
\end{subfigure}
\begin{subfigure}{.23\textwidth}
  \centering
  \includegraphics[width=1\textwidth]{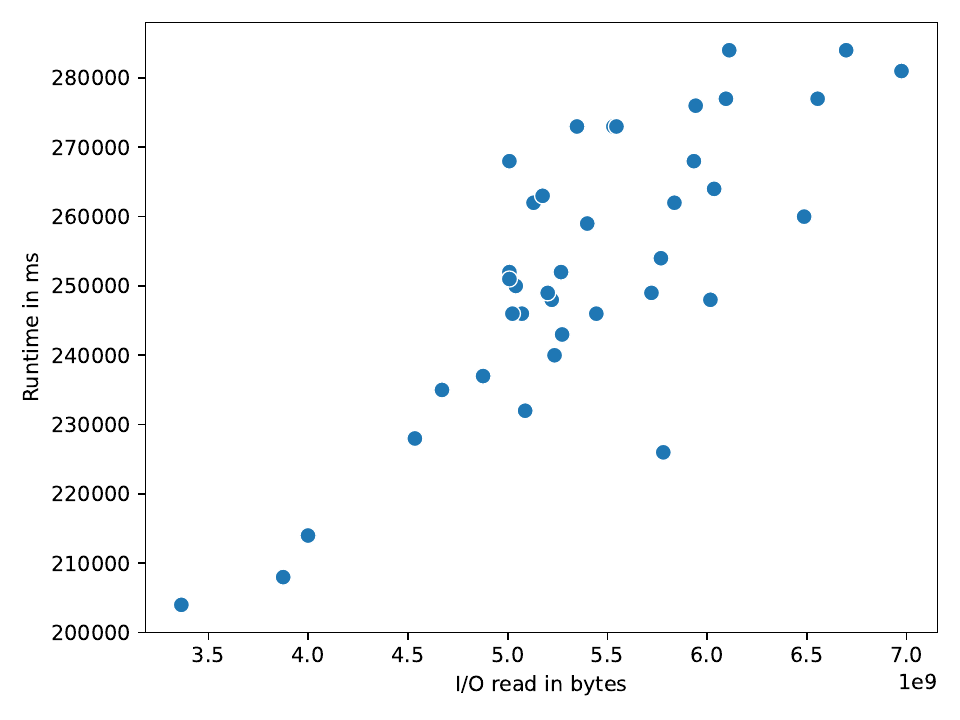}
  \caption{Star}
  \label{fig:sub-sixth}
\end{subfigure}
\begin{subfigure}{.23\textwidth}
   \centering
  \includegraphics[width=1\textwidth]{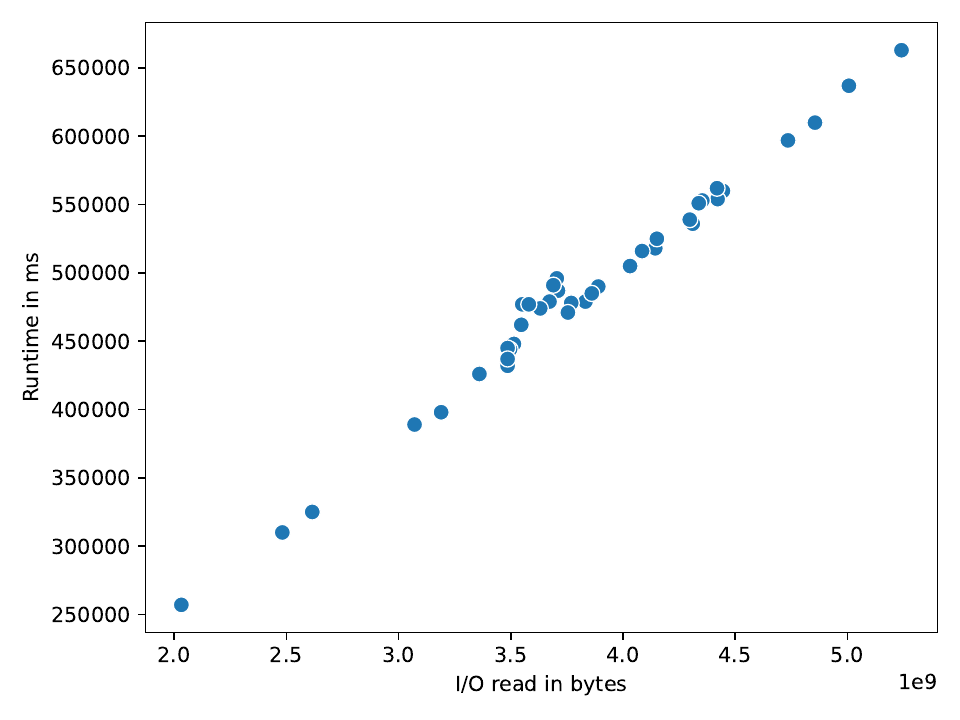}
  \caption{Picard}
  \label{fig:sub-seventh}
\end{subfigure}
\begin{subfigure}{.23\textwidth}
  \centering
  \includegraphics[width=1\textwidth]{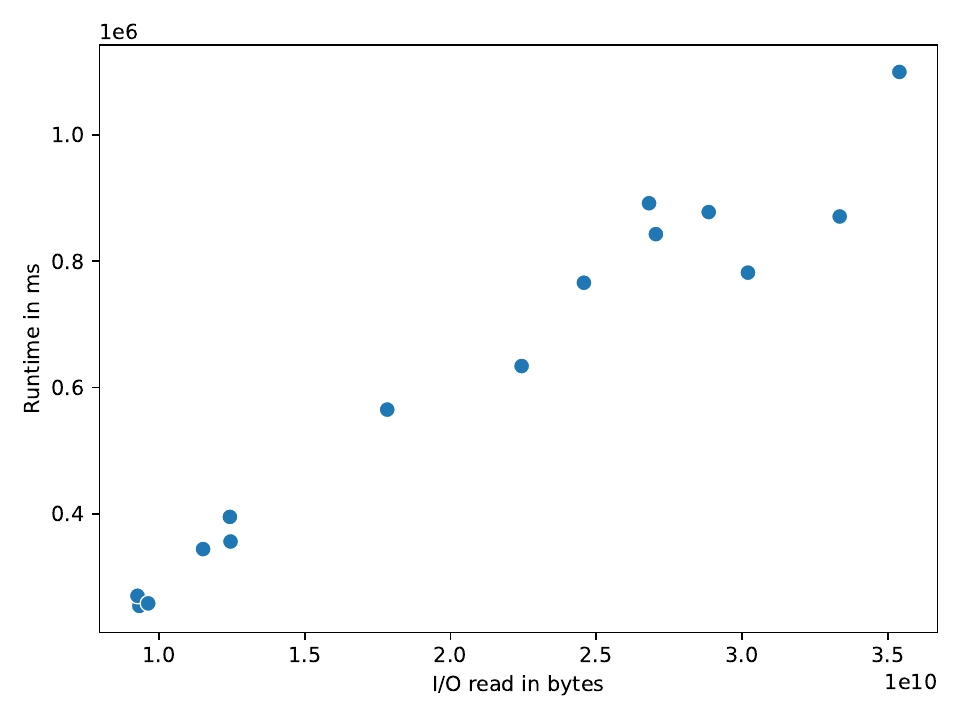}
  \caption{Bowtie2}
  \label{fig:sub-eighth}
\end{subfigure}
\begin{subfigure}{.23\textwidth}
   \centering
  \includegraphics[width=1\textwidth]{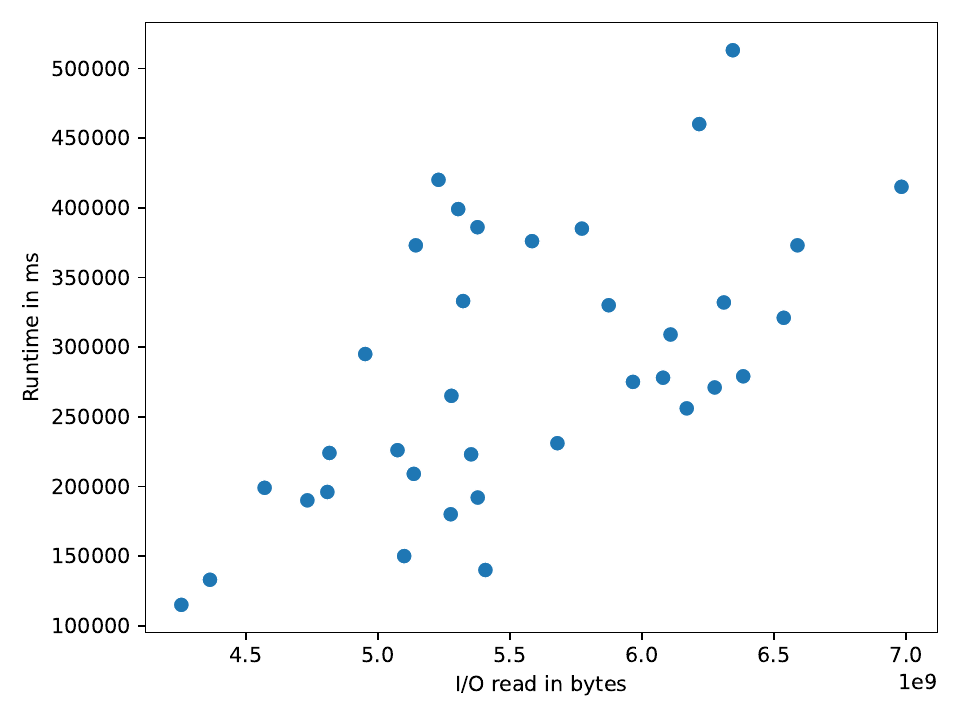}
  \caption{Fastp}
  \label{fig:sub-ninth}
\end{subfigure}
\begin{subfigure}{.23\textwidth}
  \centering
  \includegraphics[width=1\textwidth]{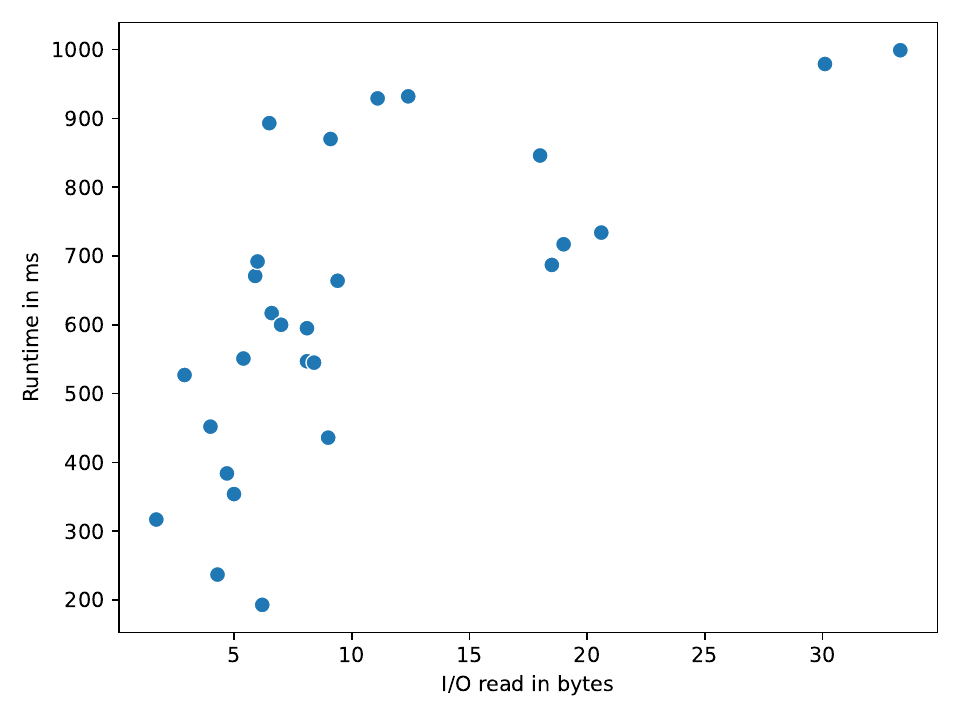}
  \caption{Gffread}
  \label{fig:sub-tenth}
\end{subfigure}
\caption{The plots show the relationship between I/O read (placeholder for data input size~\cite{da2015online}) and the task runtime. Seven tasks show a Pearson correlation coefficient of $p > 0.8$, two tasks show a $p > 0.6$, and the MultiQC task shows a Pearson correlation of $p < 0.2$.  }
\label{fig:impactSampleSize}
\end{figure}

In A5, we assume that a task's runtime scales linearly with the input data size, allowing us to use a linear task model.
This is a frequently observed pattern for big data tasks~\cite{bux2017hi,john2021evaluation,chen2013improving,ananthanarayanan2010reining}, which we also analyzed for the most popular bioinformatics tasks in Figure~\ref{fig:impactSampleSize}.
The figure shows the relationship between a task's input, the I/0 read, and a task's runtime.
Four tasks show a Pearson correlation coefficient of $p > 0.9$, which can be interpreted as very strongly correlated~\cite{schober2018correlation}. Three tasks show a $p > 0.8$, indicating a strong correlation. Two tasks exhibit a Pearson correlation coefficient of $p > 0.6$, suggesting a moderate correlation. Lastly, one task demonstrates a Pearson correlation of $p < 0.2$, representing a weak or negligible correlation.

When the scaling behavior of some workflow tasks can not be modeled linearly, i.e., A5 is violated, these tasks will yield a high prediction error.
However, while showing a high prediction error, the slope in the linear model can still describe that Task X might grow faster than Task Y, giving the resource management methods not accurate estimates but still a relative comparison of runtimes.

\subsection{Local and Target Infrastructure Profiling}
\label{subsec:approach-profiling}

\begin{table}[t]
\centering
\caption{The table shows the top ten Unix and bioinformatic tasks by their usage in all nf-core workflow repositories.}
\begin{tabular}{lr|lr}
Bioinformatic Task & Usage & Unix Task & Usage \\ \hline
multiqc            & 85\%  & echo      & 85\%  \\
fastqc             & 74\%  & sed       & 77\%  \\
samtools           & 58\%  & cat       & 77\%  \\
bedtools           & 23\%  & mkdir     & 71\%  \\
bwa                & 22\%  & ln        & 62\%  \\
STAR               & 20\%  & mv        & 57\%  \\
picard             & 20\%  & grep      & 55\%  \\
bowtie2            & 18\%  & touch     & 54\%  \\
fastp              & 16\%  & cut       & 47\%  \\
gffread            & 15\%  & gzip      & 42\% \\ 
\end{tabular}
\label{tab:nfCoreAnalysis}
\end{table}

Our method is intended to run on the scientist's local computer but can also be applied on a remote machine.
We expect that the local machine is different from the target infrastructure.  
Therefore, we conduct a short profiling phase to gather detailed infrastructure metrics and to measure machine characteristics and differences.

For unknown workflow workloads, we use general microbenchmarks that analyze the local and all target nodes' performance characteristics like CPU speeds, memory speed, and random and sequential I/O.
These microbenchmarks can be executed in parallel on all machines in the cluster and take a very short time for each node, typically less than a minute in total.

For workflows (partially) consisting of frequently used tasks or scientists willing to provide their own benchmarks, application-specific benchmarks can be executed to achieve more accurate results.

We analyzed the publicly available nf-core workflow repository~\cite{yates2021reproducible} that, at the state of writing this paper, consists of 74 real-world workflows, mainly from the bioinformatic domain, in different development statuses, from deprecated to newly created.
Table~\ref{tab:nfCoreAnalysis} shows the top ten Unix and bioinformatics tasks by their usage across workflows in the repository, e.g., the task multiqc is used in 85\% of the existing nf-core workflows.
The list of Unix tasks contains well-known applications that either do not depend on the data input at all, e.g., mkdir, or yield known patterns, e.g., gzip or grep.
The results for the most used bioinformatic tools yield that there are frequently used tools despite many different use cases in the bioinformatic domain.
Summarizing our results, many workflow tasks, at least in the same domain, are recurring, opening the space for application-specific benchmarks that can show performance differences between machines more accurately.  

The infrastructure profiling will be rerun automatically whenever a cluster's resource manager detects hardware changes.
Beyond the scope of runtime prediction, the microbenchmarks can be extended with tools that analyze network properties to consider communication aspects.

\subsection{Data Sampling and Local Workflow Execution}
\label{subsec:approach-sampling}

Our method works out of the box without any historical traces
However, to train our Bayesian regression model, training data needs to be generated a-priori.
Therefore, we want to run the workflow locally with small inputs.

To this end, we pick one of the original input files and downsample it to obtain diverse yet small (and hence fast) inputs as input for the learner. 
For instance, in remote sensing or astronomic workflows, a single image could be split into smaller ones keeping the resolution or decreasing the resolution while leaving the image section the same.
In genomics, downsampling refers to splitting one of the many input samples with millions of short sequence reads into multiple smaller partitions.
The scientist can omit the downsampling step by providing small input data partitions, e.g., satellite images with smaller resolutions or a sample with fewer reads.

Next, we take the generated inputs and run the workflow locally with them.
During this execution, monitoring data, e.g., task runtimes, input sizes, read/write I/O, are collected, which serve as the input for the prediction model.
While running the workflow locally with a large set of such partitions covering a large range of data sizes tends to improve the accuracy of the prediction model, fewer and smaller partitions can be executed faster and lead to quicker but mostly more imprecise runtime predictions.
Hence, the sizes of the samples and the number are obviously important aspects.
We studied the impact of both factors and concluded that at least three partitions with an accumulated size of at least 10\% of the downsampled input file should be used for accurate results.
For instance, if a workflow's input consists of 2,000 satellite images, each 1 GB, at least three partitions should be created with an accumulated size of at least 0.1 GB.

\subsection{Local Prediction Model Training}
\label{subsec:approach-model}

In this step, we train our Bayesian linear regression model for each task to predict the respective task's runtime.
Therefore, we use the monitoring data from the local workflow execution as data points.

According to our assumption A5, we decide to use a linear model since our experiments and many related works have shown a linear correlation between input size and runtime.~\cite{bux2017hi,john2021evaluation,chen2013improving,ananthanarayanan2010reining}. 
Before training our Bayesian linear regression model, we check for such a linear correlation between the task's input size and the task's runtime.
We argue to use the uncompressed input data size, specifically since workflows frequently operate on compressed data. 
For example, in genomic workflows, the de facto standard file format for storing biological sequences is fastq which is compressed with Gzip.
Gzip can compress larger files more efficiently, especially when dealing with repetitive data, leading to a non-linear file size increase.
In such scenarios, using the input data size as the independent variable in the linear correlation test would lead to a distorted result. 
We do this linear correlation detection first to ensure that the input file size actually impacts the task's runtime, and the relation is not a constant function.
A simple example of a task with a constant function would be the Unix command \emph{head} which is independent of its input size.
For the correlation test, we use the Pearson correlation coefficient which can measure linear relationships.
We define the relationship as strongly correlated if $p$ is greater than 0.75~\cite{schober2018correlation}.
In cases where no strong correlation can be detected, i.e., $p<0.75$,  we predict the task's median runtime, independent of the concrete input size.
Otherwise, we use a Bayesian linear regression to predict the runtime.

One of the main advantages of using the Bayesian approach is that we can train it on a small training data set~\cite{mcneish2016using,lee2004evaluation}, which is especially useful since the local profiling only delivers a few training points for each task.
Additionally, instead of predicting a point value, the Bayesian approach also yields an uncertainty value for this prediction according to a distribution.
Therefore, we also provide a lower and upper uncertainty at different confidence levels to express that the point estimate probably is not accurate.

\begin{figure}
\centering
    \includegraphics[width=1\columnwidth]{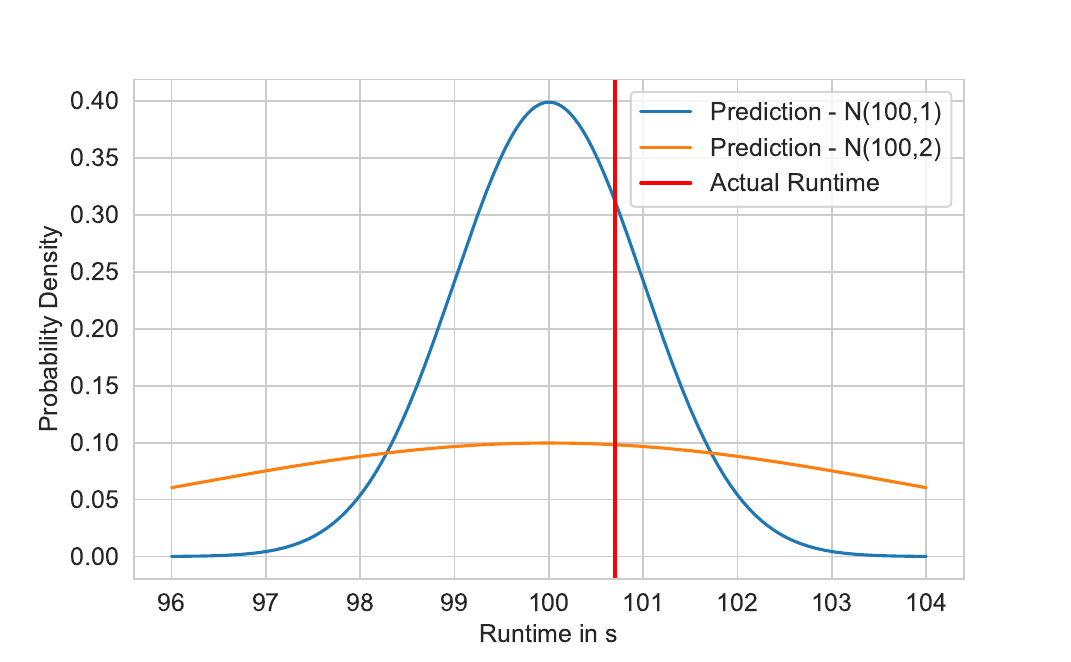}
    \caption{Posterior Prediction for a single task with two Gaussian prior distributions and the actual runtime.} 
    \label{fig:bayes_posterior}
\end{figure}
For instance, Figure~\ref{fig:bayes_posterior} shows the prediction for a single task.
The mean for both predicted values is 100 seconds, while the true runtime is 100.7 seconds.
For the prediction with the blue line, which uses a Gaussian prior with $\sigma^2=1$, the predicted value is in a confidence interval of 26\% uncertainty.
Selecting a Gaussian prior with a bigger $\sigma^2$ would flatten the curve and could express more uncertainty about the real value.
Here, the actual runtime is in a confidence interval of 14\% uncertainty.
A scheduler can consider this uncertainty and plan with it, which would not be possible for frequentist approaches.

In contrast to these frequentist approaches, we try to find a posterior distribution for our model parameters.
Specifically, our model computes a posterior distribution depending on the input values as shown in equation~\eqref{formula:bayes_1}.
We assume that $\epsilon_i$ and $y_i$ are normally distributed with $\epsilon_i \sim N(0, \sigma^2) $ and $ y_i \sim N(X\beta;\sigma^2)$

\begin{equation}
\label{formula:bayes_1}
    y_i = X\beta + \epsilon_i;
\end{equation}

Our Bayesian approach now tries to maximize the posterior

\begin{equation}
\begin{aligned}
    \max_{\beta} P(\beta|y_i);
\end{aligned}
\end{equation}

Applying Bayes theorem and a short equation transformation, we get to

\begin{equation}
\begin{aligned}
    \max_{\beta} P(y_i|\beta) P(\beta);
\end{aligned}
\end{equation}

where the first term, $P(y_i|\beta)$, the likelihood, can be computed with our previous assumption about the distribution of $y_i$. For the second term, the posterior, $P(\beta)$, we have to assume a distribution.
We decided to set the prior to a Gaussian distribution, which results in an L2 regularization for our Bayesian regression.

The gathered metrics from the local execution serve as possible features for the vector $X$, e.g., peak memory, read I/O, or CPU usage.
From the set of these features, we have to remove the features unknown in advance, e.g., the peak memory is only available after a task execution. 
Therefore, we decided to use the uncompressed input data size as a feature.
In our prediction model, $x_i$ is a scalar, i.e., the uncompressed input data, and $y_i$ is the runtime.

\subsection{Runtime Extrapolation for Target Infrastructure}
\label{subsec:approach-targetCLuster}

With the Bayesian Linear regression model created, we can now predict the tasks' runtime with arbitrary input sizes for nodes with the same hardware as the local machine.

However, our aim is to predict task runtimes for all different kinds of infrastructure nodes.
Therefore, we want to extrapolate the local runtime by calculating \mbox{$t_{target} = t_{local} \cdot f_t$}, where $t_{target}$ denotes the requested runtime on a single target machine, $t_{local}$ the runtime on the local machine, and $f_t$ the runtime factor which we need to determine.

To this end, we take the gathered data from the local and target infrastructure profiling.
In our method, we introduced general and application-specific microbenchmarks as profiling tools.
When using general microbenchmarks, tasks are treated as actual black-boxes.
Therefore, the runtime extrapolation relies on the gathered infrastructure benchmarks.
A node's I/O capabilities are essential to a task's runtime since most tasks read its input file and write some output.
Another essential factor is the CPU speed.
Therefore, we include the node's I/O and CPU characteristics in the extrapolation process. 

We decided to weight CPU and I/O equally.
With this weighting, we can now extrapolate the runtime.
We define the runtime factor $f_t$ for each task as follows:

\begin{equation}
\begin{aligned}
  f_t = 0.5 \cdot \frac{cpu_{local}}{cpu_{target}} +
        0.5 \cdot \frac{io_{local}}{io_{target}}
\end{aligned}
\label{equation::machineFactor}
\end{equation}

where the subscript \emph{local} indicates the benchmark value on the local machine and \emph{target} the benchmark value on the target machine.

For instance, the goal is to extrapolate the runtime of task T1 from the local machine to machine A1.
Our Bayesian model predicts that T1 takes 100 seconds on the local machine.
Table~\ref{tab:benchmakred_machines} shows the CPU events/s, taken as the \emph{cpu} variable, and the IOPs, taken as the \emph{io} variable.
Putting the values into equation~\eqref{equation::machineFactor} would lead to a factor of $f_t=1.7$.
Multiplying this factor with the local runtime, the task T1 is expected to run 170 seconds on A1.

Whenever application-specific benchmarks exist, we do not have to rely on the machine benchmarks.
Instead, metrics from these specific task benchmarks can be incorporated into extrapolating the runtime.
Here, the factor can be estimated by considering the benchmark value and setting them into relation:

\begin{equation}
\begin{aligned}
  f_t = \frac{val_{local}}{val_{target}};
\end{aligned}
\label{equation::machineFactorApplicationSpecific}
\end{equation}

In case application-specific benchmarks exist but not for the respective task, the runtime factor for the unknown task can be defined as the median of all existing factors \mbox{$F =\{f_{1},..., f_{n}\} $}: 

\begin{equation}
\begin{aligned}
  f_{all} = med(F).
\end{aligned}
\label{equation::machineFactorApplicationSpecificMed}
\end{equation}

\section{Implementation}
We will explain our implementation consisting of the general and the application-specific profiler, the data sampler, and the local runtime prediction module.
For reproducibility, our local predictor’s source code is provided as open source code online\footnote{github.com/CRC-FONDA/Lotaru}.

\subsection{General Infrastructure Profiler}

\begin{table*}[ht]
\centering
\caption{The results from applying the infrastructure profiling on the six different nodes.}
\begin{tabular}{c|rrrrrrrrrr}
Machine & Cores & Memory & Storage & Network & CPU events/s &  RAM score & read IOPs & write IOPs & $\sum$  Time in s \\ \hline
Local   & 8    & 16 GB     & HDD  & 1 Gbps   & 458     & 18,700     & 437       & 415  & 36      \\ 
A1   & 2 x 4    & 32 GB     & HDD  &  1 Gbps  & 223     & 11,000     & 306       & 301   & 40       \\ 
A2   & 2 x 4    & 32 GB     & HDD  & 1 Gbps  & 223      & 11,000     & 341       & 336   & 39       \\ 
N1      & 8    & 16 GB    & HDD   & 16 Gbps & 369      & 13,400     & 481       & 483    & 33      \\ 
N2      & 8    & 16 GB    & HDD  &  16 Gbps & 468      & 17,000     & 481       & 483   & 30       \\ 
C2      & 8    & 32 GB    & HDD  &  16 Gbps & 523      & 18,900     & 481       & 483   & 28      \\ 
\end{tabular}
\label{tab:benchmakred_machines}
\end{table*}

The general infrastructure profiler uses \textit{sysbench}\footnote{github.com/akopytov/sysbench} as a \mbox{microbenchmark} to measure different CPU characteristics.
\textit{Sysbench} runs a benchmark that verifies prime numbers with a limit of ten seconds and a maximum verification prime number of 20,000.
Additionally, \textit{sysbench} is used to test the memory, setting the block size buffer to one megabyte and the total memory size to 100 gigabytes.

Since we run \textit{sysbench} on computers that differ in the number of CPU cores, we decided to always set the number of benchmarked CPU threads to one.
This avoids two problems.
First, on machines with different numbers of CPU cores, a node with a few very powerful cores could score lower than a node with more but slower cores.
Second, tasks allocate a fixed number of CPU cores. 
Therefore, a benchmark value that has been obtained with more than the requested resources would be misleading.

Our profiler tests the I/O performance by using \textit{fio}\footnote{github.com/axboe/fio}.
We benchmark sequential read-write and avoid measuring random read-write characteristics since sequential access patterns prevail in data analysis tasks~\cite{cheng2015cast,vazhkudai2006constructing,vazhkudai2005freeloader}.

Note, extensive benchmarking is mandatory to examine diverse hardware design aspects~\cite{cebrian2014optimized}.
However, such extensive benchmarking is time-consuming, and the additional gathered information cannot be used to conclude task behavior on different machines since tasks are treated as black boxes.
Further, nowadays, hardware is tailored to achieve good results in popular benchmarks~\cite{puzovic2010multi,cebrian2014optimized}.
We neglect that since we want to quickly derive the relative performance differences between different nodes for the purpose of adjusting runtime predictions.

\subsection{Application-specific Profiler}
Using application-specific benchmarks, detailed information of a task's behavior on different machines can be collected. 

Application-specific benchmarks need to be executable on different machine types out of the box without installation and configuration processes.
Therefore, we use Docker containers to run the microbenchmarks isolated on the machines. 
Thus, the application-specific benchmarks are portable on different machine types.
Most popular bioinformatics tools already provide a Docker image.
We found a Docker image for all top ten bioinformatic tools from our nf-core repository analysis, such as multiqc~\cite{ewels2016multiqc}, fastqc, and samtools~\cite{li2009sequence}.
Therefore, in practice, a scientist can reuse the existing containers of their workflow tasks and run them out-of-the-box as microbenchmarks, especially since most already feature test profiles with small input data.

Recently, workflow repositories such as WFCommons~\cite{coleman2022wfcommons}, nf-core~\cite{ewels2020nf}, or the Workflow~Trace~Archive~\cite{versluis2020workflow} provide templates and data for workflow executions which can also be used for generating workflow benchmarks~\cite{Coleman2022WfBenchAG}.
Further, our repository analysis in Section~\ref{subsec:approach-profiling} showed that many workflows consist of commonly used tools, further reducing the number of tasks that need to be benchmarked.

\subsection{Data Generation for Local Prediction}

The training of our Bayesian regression models relies on the data points from locally executing the workflow with small inputs.
These inputs are either manually provided by the scientist or can be generated automatically by our downsampling method.

When generating these inputs automatically, e.g., by downsampling or slicing, the actual nature of the data needs to be considered.

Therefore, an implementation must be provided for a specific domain where a certain type of data input is used.
As all our evaluation workflows run on genome sequencing data, we implemented downsampling for genome data in the fastq format using the open-source software \textit{fastqsplitter}\footnote{github.com/LUMC/fastqsplitter} to split the inputs into partitions.
However, our method features an interface to support downsampling or slicing files in arbitrary domains.

For gathering the task runtime metrics, we choose \textit{Next\-flow}~\cite{nextflow} as a workflow management system.
We extended Nextflow's monitoring interface to collect additional data, such as compressed and uncompressed input size of tasks and the overall workflow input size.

\subsection{Prediction Interface}

We implemented Lotaru's predictor as a scientific workflow management system (SWMS) independent interface.
To support as many SWMS as possible, the input only requires a table-structured comma-separated values (CSV) file with task information.
Many workflow systems include monitoring capabilities that are able to generate such structured information by default.
Similarly, Lotaru's output is a table-structured CSV file that contains the predicted task runtimes for a workflow. 
After the task prediction process, the SWMS can read this file and assign the expected runtimes to the tasks.
Lotaru can be exported as a platform-independent executable jar.
This further increases the compatibility, allowing it to be used as an SWMS plugin and as an (online) predictor in such tools.
The collected data for training and Lotaru's outputs are persisted and can be reused to extrapolate the runtime for different cluster infrastructures.

\section{Experimental Setup}
This section describes our experimental setup, including the used infrastructure, the workflows, and the baselines.

\subsection{Infrastructure Setup and Evaluation Workflows}
\label{subsec:eval-data}

We evaluate our local prediction method for an execution on a cluster consisting of six different machines: a local machine, two machines from a heterogeneous commodity cluster, and three virtual machines in the Google Cloud Platform (GCP). 
Table~\ref{tab:benchmakred_machines} lists the machines' specifications together with the results of our microbenchmarks and their cumulative execution times.
The local machine consists of an Intel Xeon E3-1230 V2 CPU (four cores, eight threads, 3.30 GHz base frequency), 16 GB memory, an HDD,  and is connected via Ethernet with 1 Gbps.
The two machines from the commodity cluster, A1 and A2, have two Intel Xeon X5355 (four cores and 2.66 GHz base frequency) each, 32GB of memory each with different hard drives, and are connected via Ethernet with 1 Gbps.

From the Google Cloud Platform (GCP), we use N1, N2, and C2 instances as heterogeneous nodes in the cluster.
While the N1 machines are based on Intel Broadwell (8 vCPU cores and 2.00 GHz base frequency), the N2 machines use Intel Cascade Lake CPUs (8 vCPU cores and 2.80 GHz base frequency).
The C2 machines are compute-optimized and based on Intel Cascade Lake with a turbo clock of up to 3.90 GHz and 8 vCPU cores\footnote{cloud.google.com/compute/docs/machine-types}.
The machines are connected via Ethernet and provide up to 16 Gbps.

\begin{table}[t]
\centering
\caption{The used workflows, the number of input samples, their accumulated data input sizes, and the number of tasks.}
\begin{tabular}{r|rrr}
Workflow  & Input size & \# Samples & \# Tasks \\ \hline
Bacass    & 8 GB       & 4 &     5         \\
Atacseq   & 55 GB      & 12 &     14       \\
Chipseq   & 93 GB      & 6  &     14       \\
Eager     & 106 GB     & 12 &     13       \\
Methylseq & 184 GB     & 14 &    8      
\end{tabular}
\label{tab:workflows}
\end{table}

\begin{table*}[t!]
\centering
\caption{The table shows the time the respective local workflow execution takes on the scientist's machine. For our experiments, the local machine consists of an Intel Xeon E3-1230 V2 CPU with four cores and eight threads, 16 GB memory, and an HDD.}
\begin{tabular}{l|rr|rr|rr|rr|rr}
Workflow        & \multicolumn{2}{l}{Atacseq} & \multicolumn{2}{l}{Bacass} & \multicolumn{2}{l}{Chipseq} & \multicolumn{2}{l}{Eager} & \multicolumn{2}{l}{Methylseq} \\ \hline
Training Set    & 0            & 1            & 0            & 1           & 0            & 1            & 0           & 1           & 0             & 1             \\
Size in GB    & 0.43            & 0.35            & 0.04            & 0.13           & 0.36            & 0.35            & 0.13           & 0.78           & 0.52             & 0.68             \\
Time in Min & 22           & 21           & 19           & 19          & 22           & 22           & 36          & 41          & 4             & 4            
\end{tabular}
\label{tab:localWFTime}
\end{table*}

To evaluate our method, we selected five real-world bioinformatics workflows from the nf-core repository~\cite{ewels2020nf}.

The five workflows specialize in different sequence analysis: The Eager workflow~\cite{yates2021reproducible} performs analysis of ancient genomic data, Chipseq\footnote{github.com/nf-core/chipseq} analyzes Chromatin Immunoprecipitation sequencing (ChIP-seq) data, Methylseq\footnote{github.com/nf-core/methylseq} is for Bisulfite sequencing in epigenomics, Atacseq\footnote{github.com/nf-core/atacseq} for ATAC-sequencing, and Bacass\footnote{github.com/nf-core/bacass} for bacterial assembly and annotation.

Table~\ref{tab:workflows} gives an overview of these workflows, their accumulated data input sizes, the number of input sequences, and the number of tasks.
The workflow inputs are retrieved from the respective nf-core test data set.
Note that workflows often run over much larger inputs, resulting in much longer workflow runtimes for real inputs, even on large-scale clusters.

For the local workflow execution and the model training, we pick two input files for each workflow.
We downsample them to approximately 10\% of their original single input size.
For instance, the workflow Bacass contains four input files with an accumulative size of 8 GB.
Assuming all files have the same size, we downsample a single 2 GB input file to several inputs with a cumulative size of around 200MB.
We do this two times to provide two different training profiles for each workflow which are tested on the full data set.

Table~\ref{tab:localWFTime} shows the time the local workflow execution takes on our local machine.
The execution times vary between 4 and 41 minutes.
We observed that the differences in execution time between the workflows are much more considerable than comparing the execution time of a single workflow with different training sets, even though one set might have six times the size of the other (Eager-0 vs. Eager-1).
We assume that this is due to some tasks that yield a static runtime and do not scale with the input data size.
For workflows with these tasks, further downsampling would not lead to a significant decrease in local workflow execution time.

\subsection{Baselines}
\label{baselines}

Many recent workflow task prediction methods use machine-learning methods like neural networks~\cite{hilman2018task}.
Such methods require large training data sets to perform well, while we want to predict the runtime before the workflow execution and without extensive prior measurements.
Therefore, we compare our approach to three approaches that do not heavily rely on historical data to build a model and are also executable in advance, namely Online-M~\cite{da2013toward}, Online-P~\cite{da2015online}, and a Naive Approach (NA),.

The Naive Approach estimates the ratio $r_t = \frac{run_q}{d_q}$ for each training data tuple $q$ (input size $d_q$, runtime $run_q$) and takes the mean $\bar{r_t}$ for task $t$ over these ratios.
It then uses this mean ratio to predict the runtime of a task $t$ with the target input size of $d_t$, using $\bar{r_t} \cdot d_t$. 

Online-P and Online-M use density-based clustering to identify high-density areas.
Then, a cluster is determined according to the I/O read value of the task to estimate.
Since the clustering is not possible with the sparse data from the local executions, we take the data point closest to the point being estimated.
Then, a Pearson correlation between all input and output parameters is calculated.
If the data is correlated, the prediction is made based on the ratio between the output and input parameters.
If the data are uncorrelated, the Online-M method uses the mean runtime of previous task executions, while Online-P attempts to sample from a Normal or Gamma distribution first.

The three baselines use the same task training data for all experiments as Lotaru.
However, since all methods are pure predictors, they do not use the gathered microbenchmark data to fit the heterogeneous infrastructure.

We compare the three baselines to our local prediction method which we explained in detail in Section~\ref{sec:approach}.
Our evaluation provides two variations of our method, Lotaru-G, which uses the general benchmarks, and Lotaru-A, which uses the application-specific benchmarks.
Apart from the difference in benchmarking and the resulting runtime extrapolation, both variations use the same data sampling, local workflow execution, and prediction model steps.

\section{Evaluation A: Prediction Accuracy}
In Evaluation A, we evaluate the accuracy of Lotaru's task runtime predictions by comparing them to two state-of-the-art methods and a naive approach using a cluster composed of six different node types.
In the first scenario, workflows are intended to run on a homogeneous cluster consisting of a single node type. 
In this setup, the methods must predict the runtime for a single node type.
In the second scenario, the workflows are meant to execute on a heterogeneous cluster comprising six different types of nodes, thus demanding predictions for each individual node type.

\subsection{Prediction Performance for a Homogeneous Cluster}

With our first experiment, we investigate the prediction performance for a homogeneous infrastructure to get an unbiased view of the prediction models' performance.
Therefore, we assume that the target infrastructure is homogeneous and the scientist's machine is similar to the target machines.

The median prediction error (MPE) serves as our evaluation metric.
It can be calculated for every workflow or every machine type and aggregates the prediction error of the tasks inside the category.
The prediction error for an individual task $t$ is calculated as follows:

\begin{equation}
\begin{aligned}
    err_t = abs\left( \frac{predicted\_runtime - actual\_runtime}{actual\_runtime} \right). 
\end{aligned}
\end{equation}

\begin{figure}[t]
    \includegraphics[width=\columnwidth]{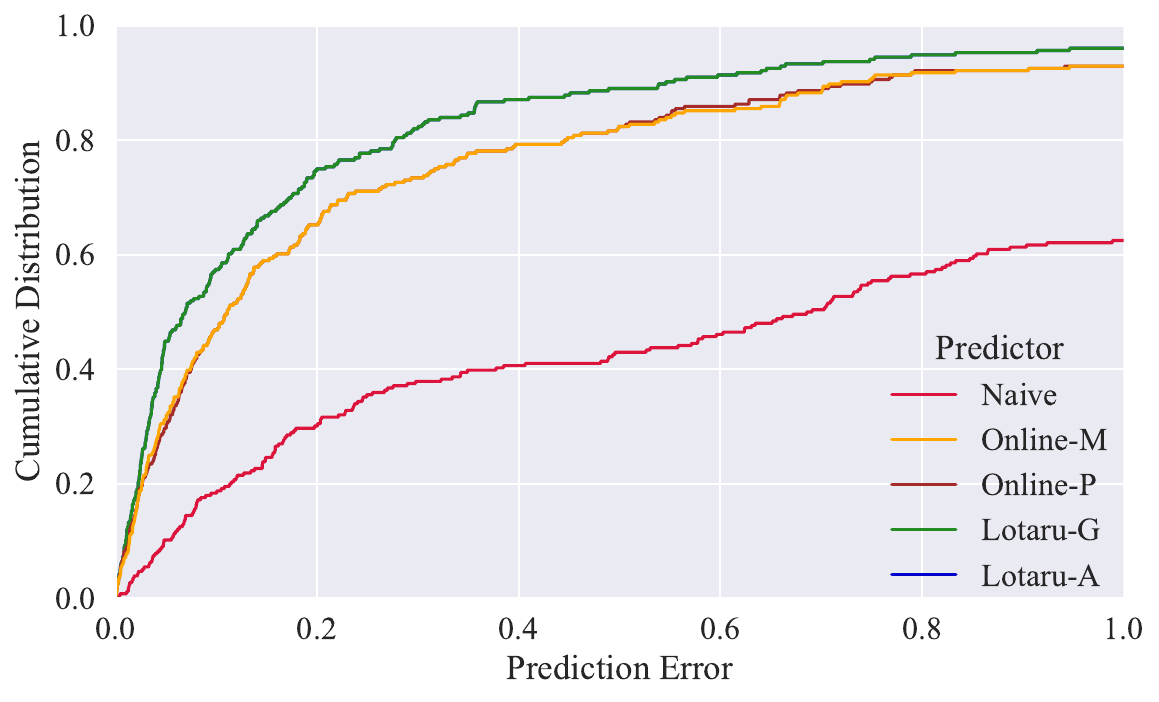}
    \caption{  CDF of the prediction error for all approaches over all workflow tasks for a homogeneous infrastructure. The green line (Lotaru-G) and the blue line (Lotaru-A) show exactly the same results, while the Orange line (Online-M) and the brown line (Online-P) are similar. For better illustration, the x-axis limit is set to 1.0, i.e., a prediction error of 100\%. The prediction error can be greater than 100\%.} \label{fig:cum_error_homogen}
\end{figure}

Figure~\ref{fig:cum_error_homogen} shows the cumulative distribution of prediction errors for all five methods over all tasks. 
Our methods, Lotaru-G and Lotaru-A, differ in the runtime extrapolation for the target machines. 
Since the infrastructure is homogeneous in this experiment, the lines are overlapping. 
Our local methods achieve the lowest prediction error and yield a median prediction error of 6.93\%. 
The baselines, Online-M and Online-P, also achieve similar results since the main difference between both approaches is the handling of uncorrelated relationships between input data size and task runtime. 
Here, Online-M estimates a runtime according to the median, whereas Online-P considers a statistical distribution. 
Both methods achieve a median error of 11.11\%, while Online-P shows slightly lower prediction errors for some tasks compared to Online-M. The Naive predictor method yields the highest median prediction error of 68.82\%.

\begin{table*}[ht]
\centering
\caption{Median prediction error (MPE) for all methods on all machines over the five experiment workflows.}
\begin{tabular}{c|rrrrr}

   & Naive   & Online-M & Online-P & Lotaru-G        & Lotaru-A        \\ \hline
A1  & 51.29\% & 40.01\%  & 40.10\%  & 17.98\%          & \textbf{16.22}\% \\ 

A2  & 50.96\% & 38.17\%  & 38.06\%  & 19.31\%          & \textbf{16.64}\% \\ 

N1  & 60.01\% & 22.87\%  & 22.87\%  & 19.06\%          & \textbf{14.60}\% \\ 

N2  & 75.12\% & 19.54\%  & 19.59\%  & 14.60\% & \textbf{13.04}\%   \\ 

C2  & 82.39\% & 28.01\%  & 28.01\%  & \textbf{15.24}\% & 15.78\%          \\ 

Median & 63.51\% & 30.87\%  & 30.69\%  & 17.33\%          & \textbf{15.18}\% \\ 
\end{tabular}
\label{tab:predHet}
\end{table*}

\subsection{Prediction Performance for a Heterogeneous Cluster}

In our second prediction experiment, we use our local machine to predict the runtimes for all target nodes A1, A2, N1, N2, and C2 for all tasks in all five evaluation workflows.

\begin{figure}[t]
    \includegraphics[width=\columnwidth]{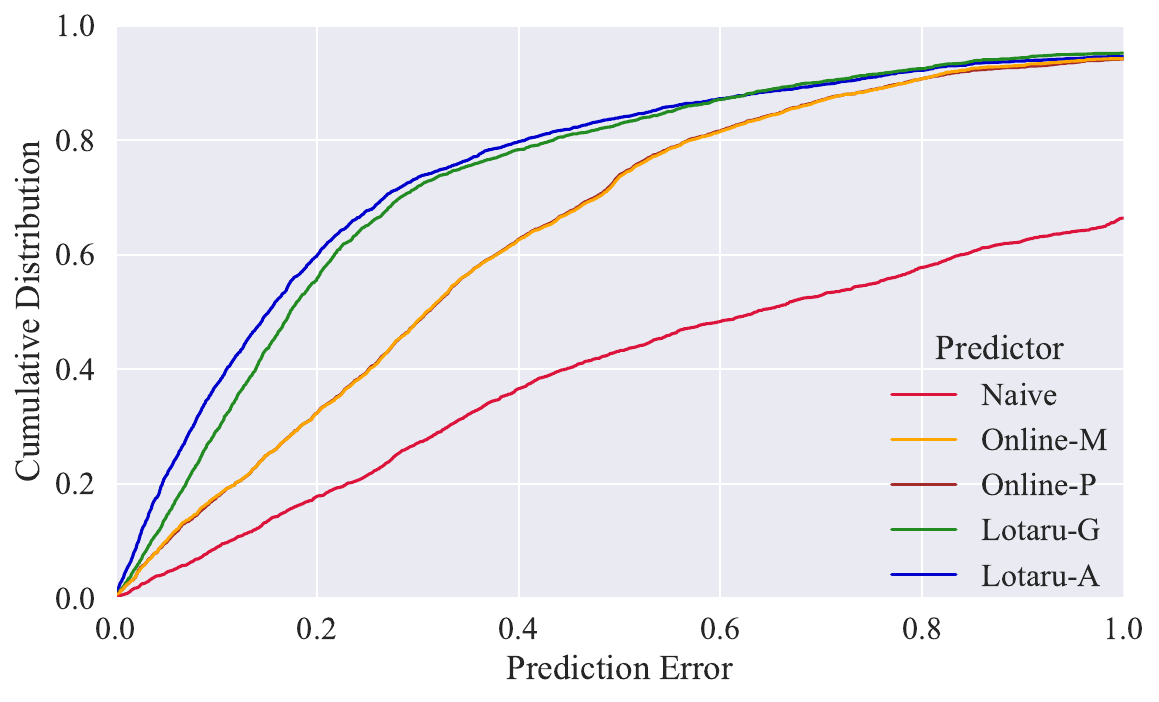}
    \caption{  CDF of the prediction error for all approaches over all workflow tasks for a heterogeneous infrastructure. The orange line (Online-M) and the brown line (Online-P) are similar. For better illustration, the x-axis limit is set to 1.0, i.e., a prediction error of 100\%. The prediction error can be greater than 100\%.} \label{fig:cum_error_heterogen}
\end{figure}

Figure~\ref{fig:cum_error_heterogen} shows the cumulative distribution of prediction errors for all task predictions on all five machines.
Our method using the application-specific benchmarks, Lotaru-A, achieves the lowest prediction errors.
The second lowest prediction errors can be observed with our method applying general benchmarks only, Lotaru-G.
For Lotaru-A 50\% of the predictions yield an error below 15.18\% and for Lotaru-G 50\% of the predictions yield an error below 17.33\%.
The best-performing baseline, Online-P, shows a median prediction error of 30.86\%.
Again, Online-M shows a similar behavior since the main difference between both approaches is handling uncorrelated relationships between input data size and task runtime. 
The Naive baseline shows the highest prediction errors.

Comparing the prediction error for a homogeneous cluster from Figure~\ref{fig:cum_error_homogen} with the prediction error for a heterogeneous cluster from Figure~\ref{fig:cum_error_homogen} shows that the cumulative distribution function (CDF) for the heterogeneous infrastructure has a lower slope compared to the CDF for the homogeneous infrastructure for each method.
Further, while the cumulative distribution of 0.5 (median) shows a difference between our local methods and Online-P of 4,18\% for homogeneous cluster, the difference increases to 13,36\% and 15,51\% for the heterogeneous cluster.

Table~\ref{tab:predHet} shows the median prediction errors for each target machine and over all machines.
For Lotaru-A, the median prediction error is in a range between 13.04\% and 16.64\%, showing nearly no correlation between the type of machine and the prediction error. 
Lotaru-G shows a lower prediction error for hardware characteristics closer to the local machine.
I.e., the prediction error for the machines N2 and C2 is lower than for A1 and A2, see also Table~\ref{tab:benchmakred_machines}.
Online-M and Online-P show a similar behavior, resulting in lower prediction errors for N2 machines.
While the differences regarding the prediction error between our method and Online-P for N1 and N2 are relatively small, they increase for A1 and A2.

\section{Evaluation B: Runtime Prediction in Resource Management}
In Evaluation B, we use the predicted runtimes from Evaluation A to investigate the impact of our achieved prediction accuracy for resource management.
Specifically, we evaluate multi-workflow scheduling, carbon-emission efficiency, and cost prediction using the established workflow simulator WorkflowSim.
To this end, we use the measurements from real-world experiments, the methods' predictions, and the machine characteristics and load them into WorkflowSim.

\subsection{Scheduling Workflows}

\begin{table*}[ht]
\centering
\caption{The table shows the deviation between the minimum makespan and the makespan achieved by the respective method over 200 different cluster configurations and all workflows. A mean deviation of 10\% translates to a 10\% longer mean makespan for the respective method compared to the best method.}
\begin{tabular}{c|rrrrrrrr}

  \multicolumn{1}{l|}{Method}        & \multicolumn{1}{c}{Mean} & \multicolumn{1}{c}{25th} & \multicolumn{1}{c}{50th} & \multicolumn{1}{c}{90th} & \multicolumn{1}{c}{95th} & \multicolumn{1}{c}{99th} & \multicolumn{1}{c}{99.9th} & \multicolumn{1}{c}{Max} \\ \hline
Naive & 72.14\% & 37.24\% & 61.20\% & 148.81\% & 158.54\% & 175.08\% & 243.27\% & 301.04\%
\\ 
Online-M  & 67.47\% & 33.05\% & 61.19\% & 148.81\% & 158.54\% & 158.54\% & 196.00\% & 292.26\%            
\\ 
Online-P  & 67.38\% & 33.05\% & 61.18\% & 148.81\% & 158.54\% & 158.54\% & 162.98\% & 250.90\% 
\\ 
Lotaru-G & \textbf{3.35}\% & \textbf{0.00}\% & \textbf{0.00}\% & \textbf{7.11}\% & \textbf{22.48}\% & \textbf{61.50}\% & 134.91\% & \textbf{195.49}\% 
\\ 
Lotaru-A & 4.79\% & \textbf{0.00}\% & \textbf{0.00}\% & 11.13\% & 33.21\% & 78.90\% & \textbf{118.83}\% & 232.93\% 
\\  \rowcolor[HTML]{EFEFEF}  
Accurate & 1.74\% & 0.00\% & 0.00\% & 3.80\% & 11.52\% & 30.56\% & 66.70\% & 120.59\%          \\ 
\end{tabular}
\label{tab:schedulingMultiple}
\end{table*}

In the scheduling experiment, we want to evaluate how well state-of-the-art scheduling algorithms can work with our predicted values.
Therefore, we feed the respective algorithm with the predicted runtime but use the actual runtime for running the tasks.
Established workflow simulation tools, such as WorkflowSim~\cite{chen2012workflowsim} or WRENCH~\cite{casanova2018wrench}, assume the presence of accurate task runtimes to execute scheduling approaches.
Therefore, we use our WorkflowSim extension WorkSim-PredError~\cite{baderReshi2022IPCCC}, which allows users to include inaccurate runtimes, e.g., the predicted runtime.
Thus, the scheduling approaches can use the predicted runtimes.

We use the state-of-the-art heuristic HEFT~\cite{heft} as the scheduling algorithm in our experiments.
In addition to the five prediction methods, we feed the HEFT scheduler with accurate runtimes as an optimum baseline.
To simulate the workflow executions on many different infrastructures, we generate 200 clusters.
Each cluster consists of 20 nodes, where the nodes are selected from the pool of previously presented cluster nodes, A1, A2, N1, N2, and C2.
The nodes' attributes, such as cores, memory, or network bandwidth, are set according to the nodes' specifications from Table~\ref{tab:benchmakred_machines}.

As described in Section~\ref{subsec:eval-data}, we applied the downsampling twice for each workflow, leading to two different training profiles for each workflow and resulting in different predictions.
Thus, we include each workflow twice with the respective predictions in the pool of possible workflows.
We always schedule two workflows from the pool of possible workflows together to create more load on the simulated cluster.

Table~\ref{tab:schedulingMultiple} shows the results for scheduling multiple workflows.
For each cluster configuration, we calculate the deviation from the minimum makespan achieved by a method.
Then, we calculate the mean and other percentile values, e.g., the 90th percentile, over these deviations.

The table shows that HEFT with accurate runtimes is able to achieve the lowest deviation values.
Noteworthy, HEFT with accurate runtimes not always leads to the lowest makespan.
Since HEFT is a heuristic, other approaches can, also randomly, sometimes find better solutions, leading to a lower makespan.
The second lowest makespan is achieved by Lotaru-G, followed by Lotaru-A.
The median performance of both methods is on par with the best method, showing a median deviation from the lowest makespan of 0.00\%.
The mean values are slightly higher than accurate runtimes, with 3.35\% for Lotaru-G and 4.79\% for Lotaru-A.
Online-P, the best-performing baseline regarding the prediction error, shows a mean makespan deviation of 67.38\% and a median makespan deviation of 61.18\%.
Online-M performs slightly worse, followed by the Naive approach, which leads to the highest makespans.
All baselines' deviations are considerably higher than our local method, translating to workflow executions taking more than 50\% longer on average.

\subsection{Carbon Efficiency}

\begin{figure}[t]
\centering
    \includegraphics[width=1\columnwidth]{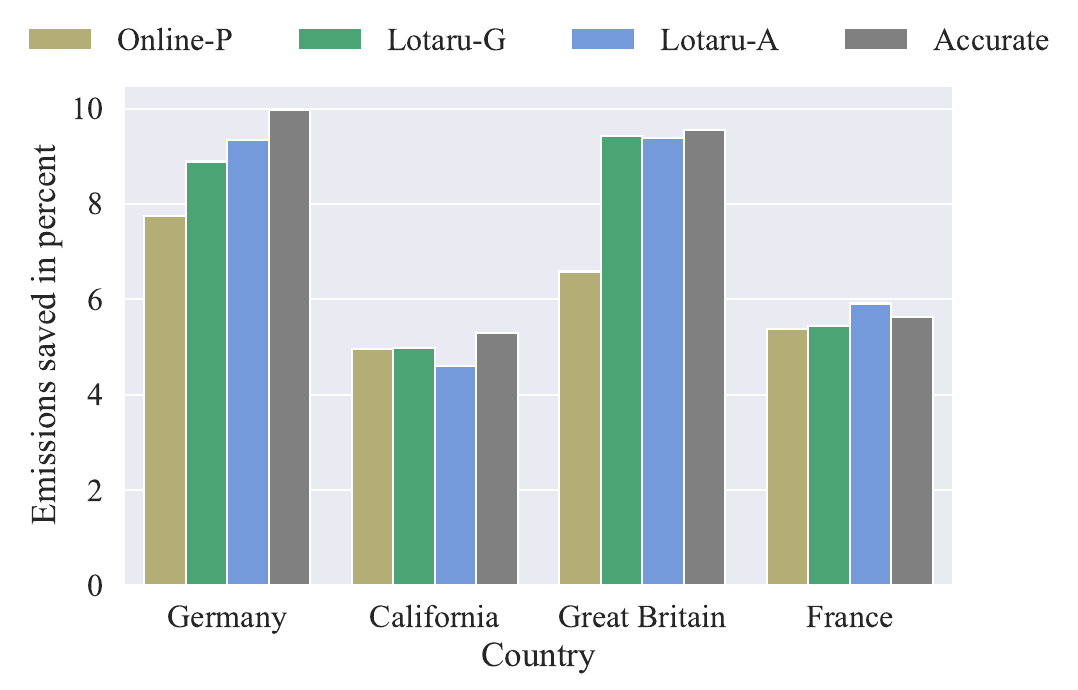}
    \caption{Carbon emission savings for semi-weekday workload shifting using runtime predictions from different methods.}
	\label{fig:workflow_shifting_semi_weekly}
\end{figure}

In this experiment, we want to examine the impact on carbon emissions when using a state-of-the-art temporal workload shifting method to reduce carbon emissions.
In Let's Wait Awhile~\cite{wiesner2021let}, workloads are shifted to times with fewer carbon emissions, e.g., when the electricity production uses renewable energy sources.
In the simulation provided by the authors, they consider a 5\% forecast error for carbon emissions but assume that runtimes estimates are accurate.
First, we extend their simulation environment and enable using predicted task runtimes.
Second, we export the schedule from all five workflows and load them into Let's Wait Awhile's simulation tool.

In the first setup, we allow shifting of tasks semi-weekly, i.e., next Monday or next Thursday at 9 am.
Figure~\ref{fig:workflow_shifting_semi_weekly} shows the emissions saved in percent by consuming energy at times with lower carbon intensity for four different regions, Germany, California, Great Britain, and France.
We include our method's predictions with general benchmarks (Lotaru-G) and application-specific benchmarks (Lotaru-A), the best-performing baseline Online-P, and accurate predictions.
In three out of four scenarios, most carbon emissions can be saved with accurate runtimes.
Only in France, Lotaru-A can save more carbon emissions.
Such behavior can occur due to the carbon-emission forecast error or prediction errors, e.g., the last workflow task yields an overprediction, resulting in an extended shifting period with lower carbon emissions. 
Over all regions, Lotaru-A is able to save the second-most carbon emissions, followed by Lotaru-G.
Using the predictions of Online-P leads to the lowest emissions saved.

\begin{figure}[t]
\centering
    \includegraphics[width=1\columnwidth]{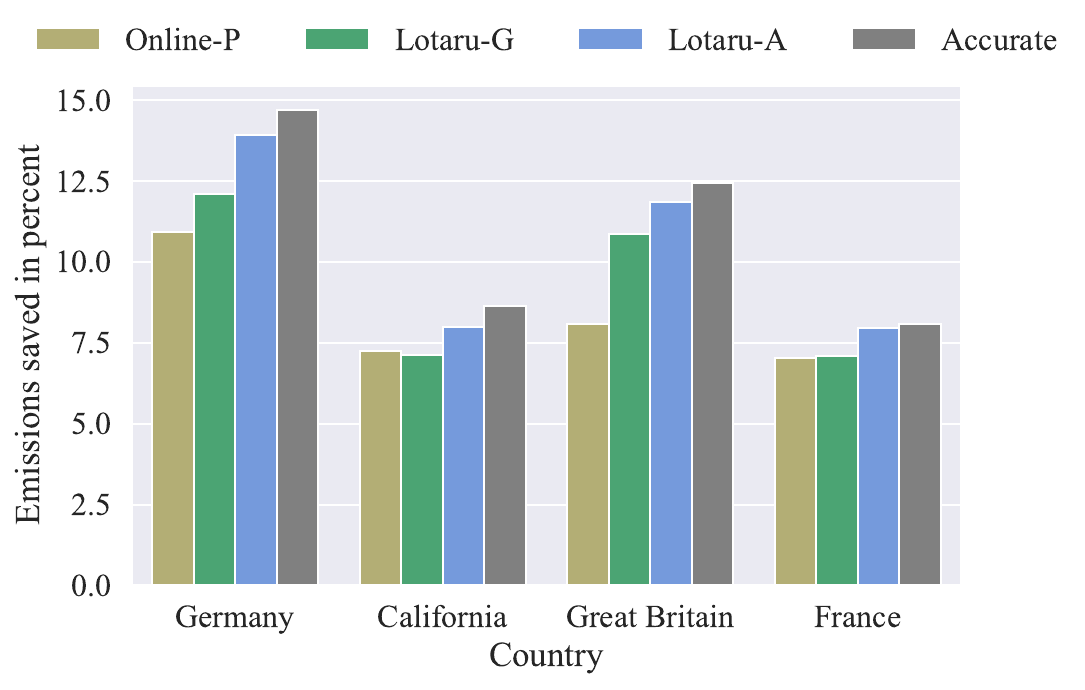}
    \caption{Carbon emission savings for next Monday workload shifting using runtime predictions from different methods.}
	\label{fig:workflow_shifting_next_monday}
\end{figure}

In a second setup, we allow task shifting to the next Monday.
Using this policy, Figure~\ref{fig:workflow_shifting_next_monday} shows that more emissions can be saved compared to shifting the workload to semi-weekly.
Depending on the region, the savings are around 0.5 times higher.
California shows the most considerable relative increase over all methods and Germany the highest percentage of emissions saved.
Contrary, Great Britain shows the smallest relative increase.
As in our previous setup, assuming accurate runtime predictions, most emissions can be saved.
Lotaru-A performs second best in all regions, followed by Lotaru-G in three out of four regions.
Again, using the predictions provided by Online-P leads to the lowest emissions savings.

\begin{table*}[t!]
    \centering
    \caption{Percentage difference between expected cost of running the workflow in the cloud and the actual cost assuming \textbf{hourly billing} of virtual machines.}
    \begin{tabular}{c|r|rrrrr}
        \multicolumn{2}{c}{} & \multicolumn{3}{c}{\normalsize\textbf{Hourly Billed - Difference}} \\
        Workflow & Training Set  & 
        Naive  &
        Online-M &
        Online-P &
        Lotaru-G &
        Lotaru-A 
        \\ \hline
			Atacseq & 0 & 38.46 & -22.03 & -22.03 & \textbf{-2.65} & -3.68 \\

			Aatacseq & 1 & 52.30 & -19.69 & -19.69 & 5.36 & \textbf{4.01} \\

			Bacass & 0 & 44.52 & \textbf{9.47} & \textbf{9.47} & 19.62 & 17.60 \\

			Bacass & 1 & 24.56 & -7.50 & -7.50 & 4.35 & \textbf{1.49} \\

			Chipseq & 0 & 57.85 & 7.84 & 7.84 & 1.93 & \textbf{-0.43} \\

			Chipseq & 1 & 61.10 & -9.78 & -9.78 & \textbf{-3.09} & -4.78 \\

			Eager & 0 & -26.97 & -27.68 & -27.97 & 10.79 & \textbf{10.33} \\

			Eager & 1 & -30.18 & -22.68 & -22.68 & 11.25 & \textbf{7.95} \\

			Methylseq & 0 & 30.95 & -12.99 & -13.14 & \textbf{2.66} & -4.27 \\

			Methylseq & 1 & \textbf{-1.53} & -11.24 & -11.24 & 4.66 & -2.14 \\
			\hline
        Median (abs) &    &   34.70  &  12.19  &  12.12 &    4.51 &  \textbf{4.14} \\
    \end{tabular}
    \label{tab:cost_cloud_hourly}
\end{table*}
\begin{table*}[t!]
    \centering
    \caption{Percentage difference between expected cost of running the workflow in the cloud and the actual cost assuming \textbf{minute-based billing} of virtual machines.}
    \begin{tabular}{c|r|rrrrr}
        \multicolumn{2}{c}{} & \multicolumn{3}{c}{\normalsize\textbf{Minute Billed - Difference}} \\
        Workflow & Training Set  & 
        Naive  &
        Online-M &
        Online-P &
        Lotaru-G &
        Lotaru-A 
         \\
        \hline
			Atacseq & 0 & 47.53 & -25.54 & -25.42 & 5.67 & \textbf{4.25} \\
			Atacseq & 1 & 60.59 & -25.19 & -24.08 & 8.09 & \textbf{7.21} \\
			Bacass & 0 & 46.00 & \textbf{8.87} & \textbf{8.87} & 21.41 & 19.79 \\
			Bacass & 1 & 25.25 & -9.57 & -9.52 & 6.42 & \textbf{4.71} \\
			Chipseq & 0 & 61.76 & 7.96 & 7.96 & 6.45 & \textbf{3.03} \\
			Chipseq & 1 & 65.01 & -10.41 & -10.32 & \textbf{-0.42} & -4.35 \\
			Eager & 0 & -39.90 & -46.38 & -46.47 & 12.47 & \textbf{11.50} \\
			Eager & 1 & -40.05 & -32.09 & -32.09 & 12.42 & \textbf{11.67} \\
			Methylseq & 0 & 36.04 & -18.70 & -18.71 & \textbf{2.90} & -5.85 \\
			Methylseq & 1 & \textbf{0.16} & -16.77 & -16.77 & 4.98 & -3.68 \\ \hline
        Median (abs) &  & 43.02   & 17.74  & 17.73 & 6.43 & \textbf{5.28} \\
    \end{tabular}
    \label{tab:cost_cloud_minute}
\end{table*}

\subsection{Cost Prediction}

Workflows are frequently executed in cloud environments~\cite{rosa2018cost,rosa2021computational,pham2017predicting,hilman2018task,dubey2018modified}.
Cloud providers offer a variety of machine configurations and charge the scientist hourly or minute-wise.
Executing long-running large-scale workflows on cloud machines can therefore be costly.
Knowing the execution cost in advance is essential but, again, requires accurate knowledge about the tasks' runtimes~\cite{rosa2018cost,rosa2021computational}.
Therefore, we want to address the problem of predicting workflow execution costs in cloud environments in our last experiment.

In WorkflowSim~\cite{chen2012workflowsim}, we simulate a cloud environment by creating a cluster with an effectively unlimited choice of virtual machines.
We use the heuristic HEFT~\cite{heft} to recommend a schedule in the cloud based on the predicted runtimes. 
Then, we integrate two setups.
In the first setup, we simulate billing on an hourly basis, e.g., Google Cloud Platform, while the second setup charges per minute, e.g., Microsoft Azure.
An overprediction in task runtime leads to higher predicted costs since it is expected that virtual machines need to be rented longer, while underpredictions lead to lower expected costs.
We expect such over- and underpredictions to be more crucial for minute-based billing.
For example, a prediction error of 15\% for a task that runs for 30 minutes has no impact on the predicted cost since the VM is billed hourly and it is not important whether the task takes 34.5 minutes or 30.  

Table~\ref{tab:cost_cloud_hourly} and Table~\ref{tab:cost_cloud_minute} show the deviation between the predicted and the actual cost of executing the workflow in the cloud.
We report the values for each workflow with both training data sets.
Positive numbers indicate an overprediction of cost, while negative numbers indicate an underprediction.
Accordingly, overpredicting costs would lead to an actual cheaper execution.
In the hourly-billed setup, Table~\ref{tab:cost_cloud_hourly}, Lotaru-A achieves the lowest deviation in six out of ten workflow executions and the lowest median deviation over all Workflows.
Lotaru-G shows a slightly higher median deviation than Lotaru-A and the lowest deviation in two out of ten workflow profiles.
The baselines Online-P and Online-M show a median deviation of 12.12\% and 12.19\%. 
This is more than 2.5 times the median deviation of Lotaru-G and nearly three times of Lotaru-A.
For minute-based billing, Table~\ref{tab:cost_cloud_minute}, the deviations for all methods, except the Naive one, are increased.
Lotaru-A shows an absolute increase of 1.14\% and Lotaru-G of 1.92\%.
The Online baselines show a much more considerable absolute increase.
In eight out of ten workflow profiles, either Lotaru-G or Lotaru-A achieve the lowest deviation.
Again, the lowest deviation from the actual cost is achieved using the Lotaru-A predictions, followed by Lotaru-G.

\section{Conclusion}
In this paper, we presented Lotaru, a method that predicts the runtime of scientific workflow tasks before executing the workflow on a heterogeneous infrastructure.
To this end, Lotaru profiles the local and target infrastructure with general or application-specific microbenchmarks, reduces the input data to quickly  profile the workflow locally, and predicts the runtime for target cluster nodes with a Bayesian linear regression based on the gathered data points from the local workflow execution and the microbenchmarks.

Our runtime estimation method is designed to be predominantly executed locally on the scientist's personal computer before a workflow is executed on a heterogeneous cluster, thus avoiding the usage of often scarce cluster resources and maximizing system efficiency.

We provide an open-source implementation of our method and an extendable interface for use with other domains that rely on different data inputs, e.g., remote sensing that uses satellite images.

In our evaluation with five real-world bioinformatics workflows on five different heterogeneous machines, we showed that Lotaru achieves low prediction errors and outperforms state-of-the-art runtime prediction baselines by a prediction error decrease of more than 12.5\%.
We further used the prediction results for several advanced resource management techniques from the literature.
Using the runtime predictions for the seminal HEFT scheduling algorithm led to the same median makespan and a mean makespan increase of less than 5\% compared to perfect a priori knowledge of task runtimes.
For carbon efficiency, again, Lotaru outperformed the baselines and showed savings close to perfect predictions.
Lastly, for cost prediction, we achieved a deviation below 5\% for hourly billing and below 6.5\% for minute-based billing, less than half of the baseline's deviations.

Our method is designed for all workflows that process multiple input files, which are, at least partly, processed separately, leading to data-parallel task executions.
For a quick local execution, either downsampling of such data needs to be possible, or the scientist needs to provide small input data.
Alternatively, different subsets of the input files can be used at the price of a longer local workflow execution phase.
Further, we assume a linear relationship between input sizes and task runtime which can be observed for many big data analysis tasks.

In the future, we plan to consider the execution on multiple distributed sites.
Further, we want to test our method with machines equipped with GPUs.
As our method additionally provides uncertainty estimates, we plan to leverage these by adjusting existing schedulers.

\section*{Acknowledgment}

Funded by the Deutsche Forschungsgemeinschaft (DFG, German Research Foundation) as FONDA (Project 414984028, SFB 1404).

\bibliographystyle{elsarticle-num}
\bibliography{references}

\end{document}